\documentclass[10pt,conference,orivec]{llncs}

\usepackage{wrapfig}
\usepackage{blindtext}
\usepackage{amsmath}
\usepackage{xcolor}
\usepackage{amssymb}
\usepackage{tikz}
\usepackage{algorithm}
\usepackage[noend]{algpseudocode}

\title{Permutation Invariance of Deep Neural Networks with ReLUs}

\author{Diganta Mukhopadhyay\inst{1}, Kumar Madhukar\inst{2}, and Mandayam Srivas\inst{3}}
\institute{
	Chennai Mathematical Institute, Chennai, India \email{digantam@cmi.ac.in}
	\and
	TCS Research, Pune, India \email{kumar.madhukar@tcs.com}
	\and
	Chennai Mathematical Institute, Chennai, India \email{mksrivas@hotmail.com}
}

\newcommand{\hide}[1]{}

\begin{document}

\maketitle

\begin{abstract}
	Consider a deep neural network (DNN) that is being used to suggest the
	direction in which an aircraft must turn to avoid a possible collision
	with an intruder aircraft. Informally, such a network is well-behaved
	if it asks the ownship to turn right (left) when an intruder approaches
	from the left (right). Consider another network that takes four inputs
	-- the cards dealt to the players in a game of contract bridge -- and
	decides which team can bid \emph{game}. Loosely speaking, if you
	exchange the hands of partners (\emph{north} and \emph{south}, or
	\emph{east} and \emph{west}), the decision would not change. However,
	it will change if, say, you exchange north's hand with east. This
	\emph{permutation invariance} property, for certain permutations at
	input and output layers, is central to the correctness and robustness
	of these networks.

	Permutation invariance of DNNs is really a two-safety property, i.e. it
	can be verified using existing techniques for safety verification of
	feed-forward neural networks (FFNNs), by composing two copies of the
	network. However, such methods do not scale as the worst-case
	complexity of FFNN verification is exponential in the size of the
	network.

	This paper proposes a sound, abstraction-based technique to establish
	permutation invariance in DNNs with ReLU as the activation function.
	The technique computes an over-approximation of the reachable states,
	and an under-approximation of the safe states, and propagates this
	information across the layers, both forward and backward. The novelty
	of our approach lies in a useful $tie$-$class$ analysis, that we introduce
	for forward propagation, and a scalable 2-polytope under-approximation
	method that escapes the exponential blow-up in the number of regions
	during backward propagation.

	An experimental comparison shows the efficiency of our algorithm over
	that of verifying permutation invariance as a two-safety property
	(using FFNN verification over two copies of the network).
\end{abstract}

\section{Introduction}
\label{sec:intro}

	Artificial neural networks are now ubiquitous. They are increasingly
	being allowed and used to handle increasingly more complex tasks, that used
	to be unimaginable for a machine to perform. This includes driving
	cars, playing games, maneuvering air traffic, recognizing speech,
	interpreting images and videos, creating art, and numerous other
	things. While this is exciting, it is crucial to understand that neural
	networks are responsible for a lot of decision making, some of which
	can have disastrous consequences if gone wrong. Thus, it is important
	to understand and verify the trustworthiness of a network, especially
	as they react with an uncertain environment. Towards this goal, this
	paper addresses the problem of verifying \emph{permutation invariance}
	of deep neural networks (DNNs).
	
	Consider a DNN that is being used to suggest the direction in which an
	aircraft must turn to avoid a possible collision with an intruder
	aircraft. Informally, such a network is well-behaved if it asks the
	ownship to turn right (left) when an intruder approaches from the left
	(right). Consider another network that takes four inputs -- the cards
	dealt to the players in a game of contract bridge -- and decides which
	team can bid \emph{game}. Loosely speaking, if you exchange the hands
	of partners (\emph{north} and \emph{south}, or \emph{east} and
	\emph{west}), the decision would not change. However, it will change
	if, say, you exchange north's hand with east. Such \emph{permutation
	invariance} properties, for certain permutations at input and output
	layers, are important to the correctness and robustness of these networks.

	Formally, given a DNN $\mathcal{N}$, permutations ${\sigma}_{in}$ and
	${\sigma}_{out}$, two vectors $B_1$ and $B_2$ of dimension as large as
	the input size of the neural network and a positive real $M$, the
	permutation invariance is defined as: \emph{if the inputs of the
	network lie between $B_1$ and $B_2$ component-wise, then permuting the
	input of the network by ${\sigma}_{in}$ leads to the output being
	permuted by ${\sigma}_{out}$ up to a tolerance of $M$. That is,}

    	\begin{equation*}
    	\begin{aligned}
		B_1 \leq x \leq B_2 \;{\Rightarrow}\; | {\sigma}_{out}(\mathcal{N}(x)) - \mathcal{N}({\sigma}_{in}(x)) | \leq M
    	\end{aligned}
    	\end{equation*}

	Permutation invariance of DNNs is really a two-safety property, i.e. it
	can be verified using existing techniques for safety verification of
	feed-forward neural networks (FFNNs), by composing two copies of the
	network. A straightforward way to do this would be encode the network
	and the property as SMT constraints, and solve it using
	Z3~\cite{DBLP:conf/tacas/MouraB08}. It is invariably more efficient,
	however, to use specially designed solvers and frameworks such as
	Reluplex~\cite{DBLP:conf/cav/KatzBDJK17} and
	Marabou~\cite{DBLP:conf/cav/KatzHIJLLSTWZDK19,DBLP:conf/sigcomm/KazakBKS19}.
	Still, these methods do not scale well, and are particularly
	inapplicable in this case (which requires doubling the network size),
	as the worst-case complexity of FFNN verification is exponential in the
	size of the input network.

	This paper proposes a technique to verify the permutation invariance
	property in DNNs\footnote{We assume the activation function to be ReLU
	(Rectified Linear Unit); written simply as `$\mathit{Relu}$'
	sometimes}. Our technique computes, at each layer, an
	over-approximation of the reachable states (moving forward from the
	input layer), and also an under-approximation of the safe states
	(moving backward from the final layer at which the property is
	specified). If the reachable states fall entirely within the safe
	region in any of the layers, the property is established. Otherwise, we
	obtain a witness to exclusion at each layer and run spuriousness checks
	to see if we can get an actual counterexample.

	The novelty of our approach lies in the way we propagate information
	across layers.  For the forward propagation of reachable states, as
	affine regions, we have introduced the notion of \emph{tie classes}.
	The purpose of tie classes is to group together the $\mathit{Relu}$
	nodes that will always get inputs of the same sign. This grouping cuts
	down on the branching required to account for active and inactive
	states of all the $\mathit{Relu}$ nodes during forward propagation.
	Intuitively, tie classes let us exploit the underlying symmetry of the
	network, with respect to the inputs and the permutation.  The backward
	propagation relies on convex polytope propagation. During the
	propagation one may have to account for multiple cases, based on the
	possible signs of the inputs to the $\mathit{Relu}$ nodes
	(corresponding to each quadrant of the space in which the polytope
	resides), leading to an exponential blow-up in the worst case. We
	address this by proposing a 2-polytope under-approximation method that
	is efficient (does not depend on LP/SMT solving), scalable, as well as
	effective. Note that the forward propagation may also be done using
	convex polytope propagation (which is how it is usually done,
	e.g.~\cite{syrenn}), but it requires computing the convex hull each
	time, which is an expensive operation.  In contrast, tie-class analysis
	helps us propagate the affine regions efficiently.

	The core contributions made in this paper are:

	\begin{itemize}
		\item an approach for verifying permutation invariance in DNNs,
			that is novel in its
			\begin{itemize}
				\item forward propagation, which is based on a useful tie-class analysis, \emph{and}
				\item backward propagation, that uses a scalable 2-polytope under-approximation method
			\end{itemize}
		\item a proof of soundness of the proposed approach,
		\item a prototype tool that implements our approach, and an
			experimental evaluation of its efficiency.
	\end{itemize}

	The rest of this paper is organized as follows. After covering the
	necessary background in Sect.~\ref{sec:prelims}, we present an informal
	overview of our approach in Sect.~\ref{sec:informal}. This is followed
	by the details of our forward and backward propagation techniques in
	Section~\ref{sec:propagate}, along with a running example that
	interleaves the two sections. We describe our implementation and
	experimental results in Sect.~\ref{sec:experiment}. The paper ends with
	a discussion of the related work (Sect.~\ref{sec:related}) and our
	concluding remarks (Sect.~\ref{sec:conc}).

\section{Preliminaries}
\label{sec:prelims}

    We represent vectors in $n$-dimensional space as row matrices, i.e., 
    with one row and $n$ columns. A linear transform $T$ from and $n$ dimensional
    space to an $m$ dimensional space can then be represented by a matrix $M$ with
    $n$ rows and $m$ columns, and we have: $T(\overrightarrow{x}) = \overrightarrow{x} M$.
        
    \subsubsection{Convex Polytopes} A convex polytope is defined as a conjunction
    of a set of linear constraints indexed by $i$ of the form $\overrightarrow{x}.\overrightarrow{v_i}
    \leq c_i$, for fixed vectors $\overrightarrow{v_i}$ and constants $c_i$. Geometrically, it is a
    convex region in space enclosed within a set of planar boundaries.  Symbolically, we can
    represent a convex polytope by arranging all the $\overrightarrow{v_i}$s into the columns of a
    matrix $M$, and letting the components of a row vector $\overrightarrow{b}$ to be constants $b_i$:
    $\overrightarrow{x} M \leq \overrightarrow{b}$.
        
    \subsubsection{Pullback} The \textit{pullback} of a convex polytope $P$ (given by
    $\overrightarrow{x} M_P \leq \overrightarrow{b_p}$), over an affine transform $T$ (given by
    $\overrightarrow{x} \;{\rightarrow}\; \overrightarrow{x} M_T + \overrightarrow{t_T}$), is
    defined as the set of all points $\overrightarrow{x}$ such that $T(\overrightarrow{x})$ lies inside
    $P$, i.e., $T(\overrightarrow{x}) \in P \Leftrightarrow \overrightarrow{x} M_T M_P \leq
    \overrightarrow{b_P} - \overrightarrow{t_T} M_P$.

    \subsubsection{Affine Region} An $n$-dimensional affine subspace is the set of all points 
    generated by linear combinations of a set of \textit{basis} vectors $\overrightarrow{v_i}, 0 \leq i
    < k$, added to a \textit{center} $\overrightarrow{c}$: $\{ \overrightarrow{x}~|~ \overrightarrow{x}
    = (\Sigma_{i=0}^{k-1} \alpha_i \overrightarrow{v_i}) + \overrightarrow{c}$, for some real $\alpha_i \}$.
    \\ We define an \emph{affine region} as a constrained affine subspace by bounding the values of
    $\alpha$ to be between $-1$ and $1$.  Formally, an affine region $A[B_A,\overrightarrow{c}]$ generated by
    a set of basis vectors $\overrightarrow{v_i}, 0 \leq i < k$, represented by a matrix $B_A$,
    is defined as the following set of points: $\overrightarrow{x} \in A  \Leftrightarrow (\exists
    \overrightarrow{\alpha} . \; \overrightarrow{x} = \overrightarrow{\alpha} B_A + \overrightarrow{c}
    \;{\wedge}\; | \overrightarrow{\alpha} | \leq 1 )$.

    \subsubsection{Pushforward} We define the \textit{pushforward} ($A_T$) of an affine region $A$
    (defined by $B_A$ and $\overrightarrow{c_A}$), across an affine transform $T$, (given by
    $\overrightarrow{x} \;{\rightarrow}\; \overrightarrow{x} M_T + \overrightarrow{t_T}$), as the set of
    points: $\overrightarrow{x} \in A_T \Leftrightarrow (\exists \overrightarrow{\alpha} . \;
    \overrightarrow{x} = \overrightarrow{\alpha} B_A M_T + \overrightarrow{c_A} M_T +
    \overrightarrow{t_T}\; , \; | \alpha | \leq 1)$. This is the image of $A$ under $T$.  (In a DNN
    context, a separate $M_T$ and $t_T$ is associated with each layer that is constructed from the
    weights and bias used at that layer.)

    \subsubsection{DNN Notation and Conventions} We number the layers of the neural
    network as $0$, $1$, $2$, and so on, upto $n-1$.  A layer is said to consist of
    an affine transform followed by a $\mathit{Relu}$ layer. The affine transform of layer $i$
    is given by $\overrightarrow{x} \;{\rightarrow}\; \overrightarrow{x} W_i + \overrightarrow{b_i}$,
    where $W_i$ are the weights and $\overrightarrow{b_i}$ are biases. We denote the input vectors by
    $\overrightarrow{x_0}$ feeding into the affine transform of layer 0, and in general for $i > 0$, the input of layer
    $i$'s affine transform (the output of the $i-1^{th}$ layer's $\mathit{Relu}$) as $\overrightarrow{x_i}$.
    The output of layer $i$'s affine transform (the input to layer $i$'s $\mathit{Relu}$) is labeled as
    $\overrightarrow{y_i}$. Finally, the output is $\overrightarrow{x_n}$. Also, we maintain copies of
    each variable's original and permuted value (using a primed notation). So, we have:

    \begin{equation*}
    \begin{aligned}
        \overrightarrow{x_0},\overrightarrow{x_0'} \;{\rightarrow}\; 
        \overrightarrow{x} W_0 + \overrightarrow{b_0} \;{\rightarrow}\; 
        \overrightarrow{y_0}, \overrightarrow{y_0'} \;{\rightarrow}\; \mathit{Relu} \;{\rightarrow}\; 
        \overrightarrow{x_1},\overrightarrow{x_1'} \;{\rightarrow}\; 
        \overrightarrow{x} W_1 + \overrightarrow{b_1} \;{\rightarrow}\;
        \overrightarrow{y_1}, \overrightarrow{y_1'} \;{\rightarrow}\; \mathit{Relu} \;{\rightarrow}\;\\
        {\cdots} \overrightarrow{y_{n-1}}, \overrightarrow{y'_{n-1}} 
        \;{\rightarrow}\; \mathit{Relu} \;{\rightarrow}\; \overrightarrow{x_n}, \overrightarrow{x_n'}
    \end{aligned}
    \end{equation*}

    Here, $W_i$ and $\overrightarrow{b_i}$ represent the action of the layer on the joint
    space of $\overrightarrow{x_i}$ and $\overrightarrow{x_i'}$. Then, the invariance property we
    wish to verify has the following form:

    \begin{equation*}
    \begin{aligned}
        B_1 \leq \overrightarrow{x_0}, \overrightarrow{x_0'} \leq B_2 \;{\wedge}\;
        \overrightarrow{x_0'} = {\sigma}_{in}(\overrightarrow{x_0}) \;{\Rightarrow}\; |
        \overrightarrow{x'_n} - {\sigma}_{out}(\overrightarrow{x_n}) | \leq M
    \end{aligned}
    \end{equation*}\linebreak

	Note that the precondition here is an affine region and the
	postcondition is a conjunction of linear inequalities, involving
	permutations.


\section{Informal Overview}
\label{sec:informal}

    Algorithm~\ref{alg:main} presents a high-level pseudocode of our approach.  The
    input to it is the network $\mathcal{N}$ with $n$ layers, and the invariance
    property given as a (\emph{pre, post}) pair of formulas. The algorithm begins
    by converting the pre-condition to an affine region by calling
    $\mathit{initPre}$ (line 3) and expressing the postcondition as a convex
    polytope by calling $\mathit{initPost}$ (line 4), without any loss of
    precision, as illustrated in the following example. Then it propagates the
    affine region forward, to obtain an over-approximation of the set of reachable
    values as an affine region at each subsequent layers (line 6). Similarly, an
    under-approximation of the safe region -- as a union of two convex polytopes --
    is calculated at each layer, propagating the information backward from the
    output layer (line 8).  If the reachable region at any layer is contained
    within the safe region, we have a proof that the property holds (lines 9-13).

    \begin{wrapfigure}[17]{l}{0.60\textwidth}
      \begin{minipage}{0.60\textwidth}
              \vspace{-0.60in}
              \begin{algorithm}[H]
    \caption{Overview of our approach}
    \label{alg:main}
    \begin{algorithmic}[1]
        \State \textbf{inputs:} $\mathcal{N}, \mathit{n}, \mathit{pre}, \mathit{post}$
        \State \textbf{globals:} reach[$n$], safe[$n$]

    \item[]

        \State reach[$0$] $\gets \mathit{initPre}(\mathit{pre}, \mathcal{N})$
        \State safe[$n-1$] $\gets \mathit{initPost}(\mathit{post}, \mathcal{N})$

        \For{$i \in [1 \ldots n)$}
        \State reach[$i$] $\gets \mathit{forwardPropagate}$(reach[$i-1$], $\mathcal{N}$)
        \EndFor

        \For{$i \in [n-2 \ldots 0)$}
        \State safe[$i$] $\gets \mathit{backwardPropagate}$(safe[$i+1$], $\mathcal{N}$)
        \EndFor

    \item[]

        \For{$i \in [1 \ldots n)$}
        \If{(reach[$i$] $\wedge$ $\neg$safe[$i$]) is unsatisfiable}
        \State \textbf{return} \textit{property holds}
        \Else \Comment{\textcolor{gray}{there must be a satisfying witness}}
        \State $\mathit{spuriousnessCheck}(\mathit{witness}, i)$
        \EndIf
        \EndFor
    \end{algorithmic}
    \end{algorithm}
    \end{minipage}
    \end{wrapfigure}

    If the inclusion check does not succeed, then our algorithm attempts to
    construct an actual counterexample from the witness to the inclusion check
    failure (see Alg.~\ref{alg:spcheck}). In general, pulling back the witness to
    the first layer is as hard as pulling back the postcondition. So, we try to
    find several individual input points that lead to something close to the
    witness at the layer where the inclusion fails, allowing us to check a number
    of potential counterexamples.  In lines 5-6 (Alg.~\ref{alg:spcheck}) we
    repeatedly apply $\mathit{pullBackCex}$ and collect these approximate pull back points
    layer by layer backwards until the input layer. We now simulate these points
    forward to check if the output of the DNN lies within the safe region in lines
    11-17. If for any point it does not, we have successfully constructed a
    counterexample.  Otherwise, if we cannot find any potential counterexamples
    (line 10), or if all the potential counterexamples are safe (line 17), the
    witness represents a spurious counterexample and the algorithm returns
    \emph{inconclusive}.

    \begin{algorithm}[h!]
    \caption{Spuriouness checking algorithm}
    \label{alg:spcheck}
    \begin{algorithmic}[1]

        \Procedure{$\mathit{spuriousnessCheck}$~}{$\mathit{counterexample}, \mathit{layer}$}

        \State $\mathit{cexes} \gets [\mathit{counterexample}$] \Comment{\textcolor{gray}{list of potential counterexamples}}
        \While{$\mathit{cexes} \neq \emptyset \vee \mathit{layer} > 0$}
                \State $\mathit{prevCexes} \gets \emptyset$
                \Statex \Comment{\textcolor{gray}{collect (approximate) pullbacks in the previous layer, for every c'example}}
                \For{$\mathit{cex} \in \mathit{cexes}$}
                    \State $\mathit{prevCexes}$ \texttt{<<} $\mathit{pullBackCex(cex, layer, \mathcal{N})}$
                \EndFor
                \State $\mathit{cexes} \gets \mathit{prevCexes}$
                \State $\mathit{layer} \gets \mathit{layer} - 1$
            \EndWhile
        \If{$\mathit{cexes} = \emptyset$}
        \State \textbf{return} \textit{inconclusive}  \Comment{\textcolor{gray}{pullback failed, no potential counterexamples}}
        \EndIf
        \For{$\mathit{cex} \in \mathit{cexes}$}
            \For{$j \in [0 \ldots n)$} \Comment{\textcolor{gray}{forward simulation of the counterexample}}
                \State $\mathit{cex} \gets \mathit{simulateLayer}(\mathit{cex}, j, \mathcal{N})$
                \If{cex $\in$ safe[$j$]} \Comment{\textcolor{gray}{spurious c'example, move on to the next one}}
                    \State \textbf{break}
                \EndIf
            \EndFor
        \State \textbf{return} (\textit{property failed}, $\mathit{cex}$) \Comment{\textcolor{gray}{actual counterexample found}}
        \EndFor
        \State \textbf{return} \textit{inconclusive} \Comment{\textcolor{gray}{all potential counterexamples are safe}}

        \EndProcedure
        
    \end{algorithmic}
    \end{algorithm}

    The details of $\mathit{forwardPropagate}$, $\mathit{backwardPropagate}$,
    $\mathit{pullBackCex}$ are discussed in the next section. In the remainder of
    this section, we present an example and describe the pre-processing part of our
    algorithm.

    \subsection{Running Example}
    \label{subsec:example-preprocess}
    
    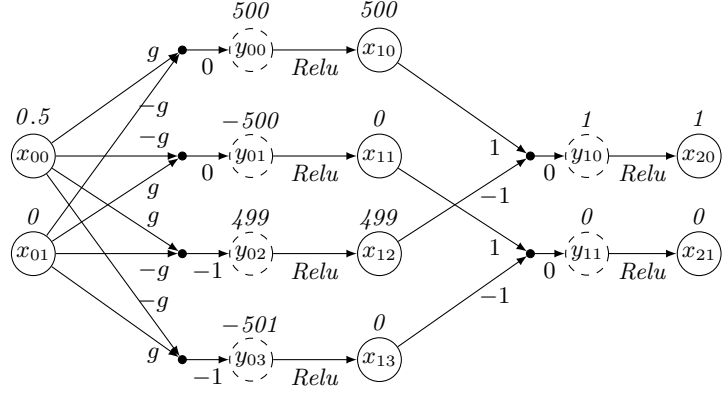
\begin{figure}
        \centering
        \begin{tikzpicture}[
        plain/.style={
            draw=none,
            fill=none,
            },
        net/.style={
            nodes={
                draw,
                circle,
                inner sep=1pt
                },
            column sep=0.4cm,
            row sep=0.2cm
            },
        pnt/.style={
            fill,
            circle,
            inner sep=0pt,
            outer sep=0pt,
            minimum size=3pt
        },
        >=latex
        ]
        \matrix[net] (mat)
        {
            |[plain]| &[1.2cm] 
                \node (p10) [pnt] {}; & 
                \node (y00) [dashed, label={[yshift=-0.1cm] $\mathit{500}$}] {$y_{00}$}; &[0.5cm] 
                \node [label={[yshift=-0.1cm] $\mathit{500}$}] (x10) {$x_{10}$}; &[1.2cm] 
                |[plain]| & |[plain]| &[0.5cm] |[plain]| \\
            \node (x00) [label={[yshift=-0.1cm] $\mathit{0.5}$}] {$x_{00}$}; &[1.2cm] 
                \node (p11) [pnt] {}; & 
                \node (y01) [dashed, label={[yshift=-0.3cm] $\mathit{-500}$}] {$y_{01}$}; &[0.5cm] 
                \node (x11) [label=$\mathit{0}$] {$x_{11}$}; &[1.2cm] \node (p30) [pnt] {}; & 
                \node (y10) [dashed, label=$\mathit{1}$] {$y_{10}$}; &[0.5cm] 
                \node (x20) [label=$\mathit{1}$] {$x_{20}$};\\
            \node (x01) [label=$\mathit{0}$] {$x_{01}$}; &[1.2cm] 
                \node (p12) [pnt] {}; & 
                \node (y02) [dashed, label={[yshift=-0.2cm] $\mathit{499}$}] {$y_{02}$}; &[0.5cm] 
                \node (x12) [label={[yshift=-0.2cm] $\mathit{499}$}] {$x_{12}$}; &[1.2cm] 
                \node (p31) [pnt] {}; & \node (y11) [dashed, label=$\mathit{0}$] {$y_{11}$}; &[0.5cm] 
                \node (x21) [label=$\mathit{0}$] {$x_{21}$};\\
            |[plain]| &[1.2cm] \node (p13) [pnt] {}; & 
                \node (y03) [dashed, label={[yshift=-0.3cm] $\mathit{-501}$}] {$y_{03}$}; &[0.5cm] 
                \node (x13) [label=$\mathit{0}$] {$x_{13}$}; &[1.2cm] 
                |[plain]| & |[plain]| &[0.5cm] |[plain]| \\
        };
        \draw[->] (x00) -- (p10) node[pos=0.8, above] {$g$};
        \draw[->] (x00) -- (p11) node[pos=0.8, above] {$-g$};
        \draw[->] (x00) -- (p12) node[pos=0.8, above] {$g$};
        \draw[->] (x00) -- (p13) node[pos=0.8, above] {$-g$};
        \draw[->] (x01) -- (p10) node[pos=0.8, below] {$-g$};
        \draw[->] (x01) -- (p11) node[pos=0.8, below] {$g$};
        \draw[->] (x01) -- (p12) node[pos=0.8, below] {$-g$};
        \draw[->] (x01) -- (p13) node[pos=0.8, below] {$g$};
        \draw[->] (p10) -- (y00) node[pos=0.5, sloped, below] {$0$};
        \draw[->] (p11) -- (y01) node[pos=0.5, sloped, below] {$0$};
        \draw[->] (p12) -- (y02) node[pos=0.5, sloped, below] {$-1$};
        \draw[->] (p13) -- (y03) node[pos=0.5, sloped, below] {$-1$};
        \draw[->] (y00) -- (x10) node[pos=0.5, sloped, below] {$\mathit{Relu}$};
        \draw[->] (y01) -- (x11) node[pos=0.5, sloped, below] {$\mathit{Relu}$};
        \draw[->] (y02) -- (x12) node[pos=0.5, sloped, below] {$\mathit{Relu}$};
        \draw[->] (y03) -- (x13) node[pos=0.5, sloped, below] {$\mathit{Relu}$};
        \draw[->] (x10) -- (p30) node[near end, below] {$1$};
        \draw[->] (x12) -- (p30) node[near end, below] {$-1$};
        \draw[->] (x11) -- (p31) node[near end, below] {$1$};
        \draw[->] (x13) -- (p31) node[near end, below] {$-1$};
        \draw[->] (p30) -- (y10) node[pos=0.5, sloped, below] {$0$};
        \draw[->] (p31) -- (y11) node[pos=0.5, sloped, below] {$0$};
        \draw[->] (y10) -- (x20) node[pos=0.5, sloped, below] {$\mathit{Relu}$};
        \draw[->] (y11) -- (x21) node[pos=0.5, sloped, below] {$\mathit{Relu}$};
        \end{tikzpicture}
        \caption{$\sigma = (0 {\rightarrow} 1, 1 {\rightarrow} 0)$, $g=1000$}
        \label{eg1}
    \end{figure}

    Consider the neural network shown in Fig.\ref{eg1}. Here, we have separated the result of
    computing the weighted sum from that of the application of the $\mathit{Relu}$ into separate
    nodes, represented by dashed and solid circles respectively. Also, we show the weights as labels
    on the arrows coming into a combination point, and biases as labels of arrows emerging from the
    point.  The arrows for weights that are $0$ have been omitted. The values at (output of) each node in the
    network for the input in the range $[0.5 \; 0]$ are shown in the diagram at that node.
    
    This network has the following symmetry property: $0 \leq x_{00}, x_{01}, x_{00}',
    x_{01}' \leq 1 \;{\wedge}\; x_{00} = x_{01}' \;{\wedge}\; x_{01} = x_{00}' \;{\Rightarrow}\;
    |[x_{40} \; x_{41}] - [x_{40}' \; x_{41}']| \leq 0.1$. This expresses the fact that flipping the
    inputs leads to the outputs being flipped.

    \subsubsection{Preprocessing:}
    
    The $W_i$ and $\overrightarrow{b_i}$ are calculated as follows: If the weights and bias
    of ayer $i$ are $W^o$ and $\overrightarrow{b^o}$, then $W_i = \bigl[\begin{smallmatrix} W^0 & 0 \\ 0 & W^0
    \end{smallmatrix}\bigr]$ and $\overrightarrow{b_i} = [\overrightarrow{b^o} ~
    \overrightarrow{b^o}]$ as we need to track both the original and permuted values at each layer. For this example we have:
    
    \begin{equation*}
    \begin{matrix}
        \begin{aligned}
            &W_0 = \\
            &\begin{bmatrix} 
                1000 & -1000 & 1000 & -1000 & 0 & 0 & 0 & 0 \\
                -1000 & 1000 & -1000 & 1000 & 0 & 0 & 0 & 0 \\
                0 & 0 & 0 & 0 & 1000 & -1000 & 1000 & -1000 \\
                0 & 0 & 0 & 0 & -1000 & 1000 & -1000 & 1000 \\
            \end{bmatrix} \\
            &\overrightarrow{b_0} = \begin{bmatrix} 0 & 0 & -1 & -1 & 0 & 0 & -1 & -1 \\
            \end{bmatrix} \\
        \end{aligned} &
        \begin{aligned}
            W_1 &= \begin{bmatrix} 
                 1 &  0 &  0 &  0 \\
                 0 &  1 &  0 &  0 \\
                -1 &  0 &  0 &  0 \\
                 0 & -1 &  0 &  0 \\
                 0 &  0 &  1 &  0 \\
                 0 &  0 &  0 &  1 \\
                 0 &  0 & -1 &  0 \\
                 0 &  0 &  0 & -1 \\
            \end{bmatrix} \\ 
            \overrightarrow{b_1} &= \begin{bmatrix} 0 & 0 & 0 & 0 \\ \end{bmatrix} \\
        \end{aligned}
    \end{matrix}
    \end{equation*}

    \subsubsection{Action of $\mathit{initPre}$ and $\mathit{initPost}$:}

    Now, $\mathit{initPre}$ calculates $\mathit{reach}[0]$ as the following affine region given by basis
    $B_0$ and center $\overrightarrow{c_0}$, and $\mathit{initPost}$ expresses $\mathit{safe}[2]$ as
    a convex polytope:
    
    \begin{equation}
    \begin{matrix}
        \begin{aligned}
            &\mathbf{reach[0]}: \\
            \exists \overrightarrow{\alpha} : [\overrightarrow{x_0} ~ \overrightarrow{x_0'}] &= \overrightarrow{\alpha} B_0 + \overrightarrow{c_0}, \;
            |\overrightarrow{\alpha}| \leq 1 \\ 
            B_0 &= \begin{bmatrix} 0.5 & 0 & 0 & 0.5 \\ 0 & 0.5 & 0.5 & 0 \\ \end{bmatrix}\\
            \overrightarrow{c_0} &= \begin{bmatrix} 0.5 & 0.5 & 0.5 & 0.5 \\ \end{bmatrix}
        \end{aligned} &
        \begin{aligned}
            &\mathbf{safe[2]}: \\
            [\overrightarrow{x_4} ~ \overrightarrow{x_4'}] \begin{bmatrix} 
                1 & 0 & -1 & 0 \\ 
                0 & 1 & 0 & -1 \\ 
                0 & -1 & 0 & 1 \\ 
                -1 & 0 & 1 & 0 \\ 
            \end{bmatrix} &\leq 
            \begin{bmatrix} 0.1 & 0.1 & 0.1 & 0.1 \end{bmatrix}
        \end{aligned}
        \label{eqn:prepost} \\
    \end{matrix}
    \end{equation}

    \subsubsection{Forward Propagation:} 
 
    $\mathit{ForwardPropagate}$ then propagates (\ref{eqn:prepost}) across the layers to get affine regions
    that are over-approximations for the reachable region for that layer. While propagation across
    the linear layer can be done easily via matrix multiplication, propagating across the $\mathit{Relu}$ layer is in general is hard,
    since we need to take into account all possible branching behaviors. We do this via a tie class
    analysis (section~\ref{subsec:forprop}) that exploits the inherent symmetry of the network and precondition. For this network,
    propagating across the first linear layer gives us an affine region given by the basis and
    center:

    \begin{equation*}
    \begin{aligned}
        B_0' &= \begin{bmatrix}  
            500 & -500 &  500 & -500 & -500 &  500 & -500 &  500 \\ 
           -500 &  500 & -500 &  500 &  500 & -500 &  500 & -500 \\ 
        \end{bmatrix} \\
        \overrightarrow{c_0'} &= \begin{bmatrix} 0 & 0 & -1 & -1 & 0 & 0 & -1 & -1 \\ \end{bmatrix} \\
    \end{aligned}
    \end{equation*}

    Then, propagating across the $\mathit{Relu}$ using the tie class analysis (section~\ref{subsec:forprop}) gives us the basis
    $B_1$  and center $\overrightarrow{c_1}$ for $\mathit{reach[1]}$. Similarly, the algorithm
    propagates across the second layer to get $B_1'$, $\overrightarrow{c_1'}$, $B_2$ and
    $\overrightarrow{c_2}$. In this case, the affine region before and after the $\mathit{Relu}$
    turn out to be the same, and there is no loss in precision going from $B_1'$ to $B_2$. The matrices are:

    \begin{equation}
    \label{eqn:eg_mats}
    \begin{aligned}
        & B_1 = &\;& B_1', B_2 = \\
        & \begin{bmatrix}  
            500 &    0 &    0 &    0 &    0 &  500 &    0 &    0 \\ 
           -500 &    0 &    0 &    0 &    0 & -500 &    0 &    0 \\ 
              0 & -500 &    0 &    0 & -500 &    0 &    0 &    0 \\ 
              0 &  500 &    0 &    0 &  500 &    0 &    0 &    0 \\ 
              0 &    0 &  500 &    0 &    0 &    0 &    0 &  500 \\ 
              0 &    0 & -500 &    0 &    0 &    0 &    0 & -500 \\ 
              0 &    0 &    0 & -500 &    0 &    0 & -500 &    0 \\ 
              0 &    0 &    0 &  500 &    0 &    0 &  500 &    0 \\ 
        \end{bmatrix} &\;&
        \begin{bmatrix} 
             500 &     0 &     0 &  -500 \\ 
            -500 &     0 &     0 &   500 \\ 
               0 &  -500 &  -500 &     0 \\ 
               0 &   500 &   500 &     0 \\ 
            -500 &     0 &     0 &  -500 \\ 
             500 &     0 &     0 &   500 \\ 
               0 &   500 &   500 &     0 \\ 
               0 &  -500 &  -500 &     0 \\ 
        \end{bmatrix}\\
        & \overrightarrow{c_1} = \begin{bmatrix} 0 & 0 & 0 & 0 & 0 & 0 & 0 & 0 \\ \end{bmatrix} &\;&
        \overrightarrow{c_1'}, \overrightarrow{c_2} = \begin{bmatrix} 0 & 0 & 0 & 0 \\ \end{bmatrix} \\
    \end{aligned}
    \end{equation}

    \subsubsection{Inclusion check:}
    
    Now, we see that if we substitute $\overrightarrow{x}$ with the form given in $\mathit{reach}[2]$ into
    $\mathit{safe}[2]$ (which is the postcondition $|[x_{40} \; x_{41}] - [x_{40}' \; x_{41}']| \leq 0.1$),
    the right side of the inequality simplifies to $0$. So, $\mathit{reach}[2]$ is included in
    $\mathit{safe}[2]$.  This is done by an algorithm (section~\ref{subsec:incl}) that checks this using an LP  solver, and since
    it succeeds in this case, it returns \textit{property holds}. 
    
    Note that for this example, it was unnecessary to perform any back propagation of the
    $\mathit{safe}[i]$ to previous layers, as the inclusion check succeeded at the output layer.
    In general, back propagation (section~\ref{subsec:backprop}) would be performed to compute under-approximations.
    Spuriousness check (section~\ref{subsec:spur}) will be needed if the inclusion check fails.

\section{Forward and Backward Propagation}
\label{sec:propagate}

    \subsection{Preparation}
    \label{subsec:prepare}

    \subsubsection{$\mathit{InitPre}$ and construction of $\mathit{reach[0]}$}:
        An input precondition on an input layer
        of the form $ \overrightarrow{B_1} \leq [\overrightarrow{x_0},\overrightarrow{x_0'}] \leq
        \overrightarrow{B_2}$ with $\overrightarrow{x_0'} = \sigma(\overrightarrow{x_0})$ can always be converted into an equivalent affine region characterized by the formula: \\
        $(\exists \overrightarrow{\alpha} :\; [\overrightarrow{x_0} \; \overrightarrow{x_0'}] =
            \overrightarrow{\alpha} V + \overrightarrow{c}, \; | \overrightarrow{\alpha} | \leq 1)$,
        with the centre as $\overrightarrow{c} = \frac{\overrightarrow{B_1} + \overrightarrow{B_2}}{2},$ and the matrix $V$ representing the set of basis vectors defined as follows. 
       
      For input size of $n$, $V$ is a concatenation of two $n$~X~$n$ matrices $[V_1 ; V_2]$ with $2n$ columns.
       The elements of $V_1$ are all 0 except the main left to right diagonal being set to $1/2(B_2[i]- B_1[i])$.
       $V_2$ is constructed by applying $\sigma$ permutation for each of the columns of $V_1$.
       
       Consider the precondition $[1, 3, 2, 2, 1, 3] \leq [x_0, x_1, x_2, x_2, x_0, x_1] \leq [2, 4, 3, 3, 4, 2]$ on a input vector of size 3 and a permutation of it.
       The matrix $V$ to express it as an affine region is
      $   
        \begin{bmatrix} 
                1/2 & 0 & 0 & 0 & 1/2 & 0  \\
                0 & 1/2 & 0 & 0 & 0 & 1/2 \\
                0 & 0 & 1/2 & 1/2 & 0 & 0  \\
            \end{bmatrix}
        $.

    \subsubsection{$\mathit{InitPost}$ and construction of $\mathit{safe_n}$:} A postcondition of the form $| \overrightarrow{x'_n} -
        {\sigma}_{out}(\overrightarrow{x_n}) | \leq M$ can be written as  $ |
        [\overrightarrow{x_n} \; \overrightarrow{x'_n}] L | \leq M$ where each column of $L$ calculates
        one of the differences of corresponding components. This can be transformed into a conjunction of two sets of
        linear constraints given by $[\overrightarrow{x_n} \; \overrightarrow{x'_n}] L \leq M$  and $
        - [\overrightarrow{x_n} \; \overrightarrow{x'_n}] L \leq -M$, which represents a 
        single convex polytope over $2n$ variables.

    \subsection{Forward Propagation using Tie Classes}
    \label{subsec:forprop}

        Let $\mathit{reach_{j}} = \{ [\overrightarrow{x_j} \; \overrightarrow{x_j'}]~|~ \exists
        \overrightarrow{\alpha} :\; [\overrightarrow{x_j} \; \overrightarrow{x_j'}] =
        \overrightarrow{\alpha} B_j + \overrightarrow{c_j}, \; | \overrightarrow{\alpha} | \leq 1\}$, be
        the affine region representing an over-approximation of reachable points at the input to layer
        $j$; $forwardPropagate$ constructs $\mathit{reach_{j+1}}$ as an affine region that is an
        over-approximation for the set of all points produced when $\mathit{reach_{j}}$ is propagated to the input
        of layer $j+1$.
    \hide{
        \begin{equation*}
        \begin{aligned}
            \exists \overrightarrow{\alpha} :\; [\overrightarrow{y_j} \; \overrightarrow{y_j'}] =
            \overrightarrow{\alpha} B_j + \overrightarrow{c_j}, \; | \overrightarrow{\alpha} | \leq 1
        \end{aligned}
        \end{equation*}
    }
        $\mathit{reach_{j+1}}$, is constructed by forward propagating $reach_{j}$ first across the
        affine transform at $j$ to produce an affine region $A_j$, which is then further forward
        propagated across the $\mathit{Relu}$ layer.
     
        Forward propagation across the linear transform given by $\overrightarrow{x} \;{\rightarrow}\;
        \overrightarrow{x} W_j + \overrightarrow{b_j}$ is straightforward and precise as it can be
        computed as a simple linear pushforward across $W_j$, i.e.,
        $A_j([\overrightarrow{y_j} \; \overrightarrow{y_j'}])   \Leftrightarrow (\exists
        \overrightarrow{\alpha} :\; [\overrightarrow{y_j} \; \overrightarrow{y_j'}] =
        \overrightarrow{\alpha} B_j' + \overrightarrow{c_j'}, \; | \overrightarrow{\alpha} | \leq 1)$,
        where $B_j' = B_j W_j$ and $\overrightarrow{c_j'}$ as $\overrightarrow{c_j} W_j +
        \overrightarrow{b_j}$.

        Propagating $A_j$ across $\mathit{Relu}$ is more complex and challenging as
        it requires, in general, a detailed case analysis of the polarity and
        strength of the components of the basis vectors and the scaling $\alpha$;
        rather than performing it precisely, $\mathit{reach_{j+1}}$ is constructed
        as an affine region over-approximates the $\mathit{Relu}$ image. Several
        methods can be used to construct an over-approximation that make different
        tradeoffs between precision and efficiency. One can construct the smallest
        affine region (or polytope) that includes all the reachable values possible
        across the $\mathit{Relu}$~\cite{syrenn}. Computing the smallest region can
        be inefficient as it is an optimization problem requiring several expensive
        LP or convex-hull calls. Our method efficiently constructs an
        over-approximate affine region that, while sub-optimal, is effective for
        checking permutation invariance properties.
        
        Our method to construct the over-approximate affine region relies on looking for similarities
        in the polarity of the components of the vectors belonging to $\mathit{reach_{j}}$ that are
        preserved when a $\mathit{Relu}$ is applied to the region. For this purpose we introduce a
        new concept called \textit{tie classes} associated with an affine region which is described
        below. \\[0.20cm]

        \noindent
        \textbf{Propagating over Relu with Tie Classes} \\[0.15cm]
        \noindent
        Given an affine region $A$ defined by a basis $\overrightarrow{v_i}$ and center
        $\overrightarrow{c}$ we define a binary relation, \emph{tied}, over the set of indices \footnote{We
        assume the indices range from 0 to $n-1$ for vectors of size $n$.} denoting the components of
        any vector $\overrightarrow{x}$ in  $A$ as follows.
        
        \begin{definition}[Tied] Given an affine region $A$ characterized by the condition $\exists
            \alpha_i :\; \overrightarrow{x} = \sum_i \alpha_i \overrightarrow{v_i} + \overrightarrow{c},
            \; | \alpha_i | \leq 1$, and two indices $i_1$ and $i_2$ in the index set, we say $i_1$ and
            $i_2$ are \textit{tied} iff for every vector $\overrightarrow{x}$ in $A$ the components at
            $i_1$ and $i_2$ have the same sign.
           \end{definition}

        It is easy to see that the binary relation being tied is an equivalence relation on the index
        set of vectors $\overrightarrow{x}$ that generates an equivalence class defined as follows.
     
        \begin{definition}[Tie Class] A \textit{tie class} for an affine region $A$ is the equivalence
            class (partitioning) of the index set for the vectors in $A$ induced by  the equivalence
            relation tied for $A$.
        \end{definition}
        Consider the affine region generated by the basis $\overrightarrow{v_i}$ and $\overrightarrow{c}$: 
        $   \overrightarrow{v_0} = [ 1   \; 0 \; 0   \; 2 ] ,\; 
            \overrightarrow{v_1} = [ 0   \; 1 \; 0.5 \; 0 ] ,\; 
            \overrightarrow{c  } = [ 0.5 \; 2 \; 1   \; 1 ]$.
            For this region, the indices 0 and 3 are tied because every vector in the region
        the component  $3$ is always $2$ times the component $0$ , since
        the component $3$ of the $\overrightarrow{v_i}$ and the $\overrightarrow{c}$ are $2$ times the
        component $0$.  Similarly, indices $1$ and $2$ are tied as well.
        For this region, the tie equivalence class is $\{ 0:\{ 0, 3\}, 1:\{1, 2 \}\}$ \\[0.2cm]
         \noindent
        \textbf{Tie class based transformation of Basis Vectors} \\[0.15cm]
        \noindent
        To help construct the basis vectors for the over-approximation of the output of $\mathit{Relu}$,
        we define a transformation of the set of basis vectors at the input to $\mathit{Relu}$.
        For each tie class $j$ in the equivalence class induced, and each vector $\overrightarrow{v_i}$
        in the input basis set, we construct a vector $\overrightarrow{v'^j_i}$ by setting all the
        components of $\overrightarrow{v_i}$ that are not in the tie class $j$ to $0$. Similarly, we get
        a $\overrightarrow{c^j}$ from $\overrightarrow{c}$ for each tie class $j$. For the example
        above, we have:

        \begin{equation*}
        \begin{aligned}
            \overrightarrow{v'^0_0} &= [ 1   & & 0 & & 0   & & 2 ] &
            \overrightarrow{v'^1_0} &= [ 0   & & 0 & & 0   & & 0 ] \\
            \overrightarrow{v'^0_1} &= [ 0   & & 0 & & 0   & & 0 ] &
            \overrightarrow{v'^1_1} &= [ 0   & & 1 & & 0.5 & & 0 ] \\
            \overrightarrow{c^0   } &= [ 0.5 & & 0 & & 0   & & 1 ] &
            \overrightarrow{c^1   } &= [ 0   & & 2 & & 1   & & 0 ]
        \end{aligned}
        \end{equation*}

        We now state the following lemma:

        \begin{lemma}\label{l1} Given $\overrightarrow{x} = \sum_i \alpha_i \overrightarrow{v_i} +
            \overrightarrow{c}$, we can write $\mathit{Relu}(\overrightarrow{x}) = \sum_{i,j} \alpha'^j_i
            \overrightarrow{v'^j_i} + \sum_j \beta_j \overrightarrow{c_j}$ where each $\alpha'^j_i$ is
            either $\alpha_i$ or is $0$, and each $\beta_j$ is either $0$ or $1$. Moreover, the components
            of $\mathit{Relu}(\overrightarrow{x})$ with indices in a tie class $j$ are $0$ if and only if
            $\alpha'^j_i$ and $\beta_j$ are $0$.
        \end{lemma}
       
        The proofs have been moved to Appendix~\ref{app:proofs} for lack of space.

        This lemma states that there exists an oracle that, given an $\overrightarrow{x}$ in
        $\mathit{reach_{j}}$, can determine whether to set each $\overrightarrow{\alpha'^j_i}$ to
        $\overrightarrow{\alpha_i}$ or $0$ and each $\beta_j$ to $0$ or $1$  so that we can express
        $\mathit{Relu}(\overrightarrow{x})$ in the above form. Regardless of what the oracle chooses we
        can always replace the condition $\overrightarrow{\alpha'^j_i} = \overrightarrow{\alpha_i}
        \;{\vee}\; \overrightarrow{\alpha'^j_i} = 0$ with $|\overrightarrow{\alpha'^j_i}| \leq 1$ as an
        over-approximation. Now, if we can somehow replace $\sum_j \beta_j \overrightarrow{c_j}$ with a
        single vector, we will have found our output affine region. The following theorem proves that we
        can replace this sum with $\mathit{Relu}(\overrightarrow{c})$.
        
        \begin{theorem}\label{t1} Given $\overrightarrow{x} = \sum_i \alpha_i \overrightarrow{v_i}
            + \overrightarrow{c}, |\overrightarrow{\alpha_i}| \leq 1$, in an affine region $A$, there are scalars
            $\alpha'^j_i$ such that:\\
            \indent
                1. $\mathit{Relu}(\overrightarrow{x}) = \sum_{i,j} \alpha'^j_i \overrightarrow{v'^j_i} +
                    \mathit{Relu}(\overrightarrow{c})$\\
            \indent
                2. $|\overrightarrow{\alpha'^j_i}| \leq 1$ for all $i$ and $j$.
        \end{theorem}

        The above theorem proves that the affine region given by the basis $\overrightarrow{v'^j_i}$ and
        the center $\mathit{Relu}(\overrightarrow{c})$ is an over-approximation for the Relu image of $A$. 
        Given $\overrightarrow{v_i}$ and $\overrightarrow{c}$, it is easy to
        compute $\overrightarrow{v'^j_i}$ and $\mathit{Relu}(\overrightarrow{c})$ if we know what the
        tie classes are, since this only involves setting certain components to $0$.  All we need to do
        now is compute the tie classes for the given $\overrightarrow{v_i}$ and $\overrightarrow{c}$.
      
        \begin{wrapfigure}[13]{l}{0.60\textwidth}
          \begin{minipage}{0.60\textwidth}
            \vspace{-0.60in}
            \begin{algorithm}[H]
            \caption{Checking tiedness}
            \label{alg:tcheck}
            \begin{algorithmic}[1]
                \State \textbf{inputs:} $A, \overrightarrow{v_i}, \overrightarrow{c}, i_1, i_2$
                
                \If{$\forall j : \frac{v^{i_1}_j}{v^{i_2}_j} = \frac{c^{i_1}}{c^{i_2}}$} \textbf{return} \textit{tied}
                \ElsIf{ $\overrightarrow{c_{i_1}} \geq 0$ and $\overrightarrow{c_{i_2}} \geq 0$ }
                \If{$i_1$ or $i_2$ component of some $\overrightarrow{x} \in A$ $< 0$} \textbf{return} \textit{not tied}
                    \Else ~\textbf{return} \textit{tied}
                    \EndIf
                \ElsIf{ $\overrightarrow{c_{i_1}} < 0$ and $\overrightarrow{c_{i_2}} < 0$ }
                    \If{$i_1$ or $i_2$ component of some $\overrightarrow{x} \in A$ $> 0$} \textbf{return} \textit{not tied}
                    \Else ~\textbf{return} \textit{tied}
                    \EndIf
                \Else ~\textbf{return} \textit{not tied}
                \EndIf
                    
            \end{algorithmic}
        \end{algorithm}
        \end{minipage}
        \end{wrapfigure}

        \subsubsection{Computing Tie Classes} To compute tie classes, for every
        pair of indices $i_1$ and $i_2$, we check whether $i_1$ and $i_2$ are tied,
        and then group them together. One way to check if two $i_1$ and $i_2$ are
        in the same tie class using two LP queries involving the $\alpha_i$: one
        which constraints the value of component $i_1$ of $x$ to positive and
        component $i_2$ to negative, and vice versa. If any of these are feasible,
        $i_1$ and $i_2$ cannot be in the same tie class. Else, they are in the same
        tie class. This needs to be repeated for each pair of $i_1$ and $i_2$,
        which amounts to $n*(n-1)$ LP calls for $n$ $\mathit{Relu}$ nodes, which is
        inefficient. Instead, we state another property of tie classes that will
        allow us to compute the tie classes more efficiently:
        
        \begin{theorem}\label{t2} Two indices $i_1$ and $i_2$ are in the same tie class if and only if one
            of the following is true:\\
            \indent
                1. The $i_1$ and $i_2$ components of $\overrightarrow{x}$ are always both positive.\\
            \indent
                2. The $i_1$ and $i_2$ components of $\overrightarrow{x}$ are always both negative.\\
            \indent
                3. The vector formed by the $i_1$ and $i_2$ components of the $\overrightarrow{v_k}$ and
                    $\overrightarrow{c}$ are parallel. In other words, if $v^l_k$ is the $l$-component
                    of $\overrightarrow{v_k}$, and $c^l$ is the $l$ component of $\overrightarrow{c}$,
                    then $[ v^{i_1}_1, v^{i_1}_2, {\cdots} c^{i_1} ] = k[ v^{i_2}_1, v^{i_2}_2, {\cdots}
                    c^{i_2} ]$ for some real $k > 0$. 
        \end{theorem}

        Algorithm \ref{alg:tcheck} uses Theorem \ref{t2} to check if $i_1$ and $i_2$ are in the same tie
        class. The queries in lines 5, 7, 12 and 14 can be reduced to looking for $\alpha_j$ such that
        $\sum_j \alpha_j v^{i_1}_j + c^{i_1} < 0$. Such queries can be solved via an LP call, but we use
        Lemma \ref{l5} to avoid LP calls and check these queries efficiently.
       
        \begin{lemma}\label{l5} The maximum and minimum values of $ \sum_i \alpha_i v_i $, for real
            $\alpha_i$, fixed real $v_i$, constrained by $| \alpha_i | \leq 1$, are $\sum_i |v_i|$ and
            $-\sum_i |v_i|$ respectively.    
        \end{lemma}

        If the network has a lot of inherent symmetry with respect to the input permutation, it is more
        likely for different neurons in the same layer to be tied together, leading to larger tie
        classes. This, in turn, reduces the number of basis vectors required to construct our
        over-approximation of the $\mathit{Relu}$ image, and improves the quality of the
        over-approximation. Thus, we can expect our over-approximation to perform well for checking
        permutation invariance.

    \subsection{Backward (Polytope) Propagation}
    \label{subsec:backprop}

        The goal of backward propagation is, given a convex $P: \overrightarrow{x} L \leq
        \overrightarrow{u}$, to symbolically construct a region, that is a reasonable
        under-approximation of \break$\mathit{WeakestPrecond}(\mathit{Layer}, P)$. Back propagating $P$ across the linear part
        of a layer is easy as it can be done precisely by simply pulling back $P$ across a linear
        transform using matrix multiplication.

        Back propagating it across $\mathit{Relu}$ is more challenging because the \break
        $\mathit{WeakestPrecond}(\mathit{Relu}, P)$ may potentially involve many quadrants, the number of which
        are worst-case exponential in the dimension of the space.
        Keeping track of all of the quadrants is infeasible. A sound single polytope solution is to use $P \land
        \overrightarrow{x} \geq 0$ ignoring the entire ``non-positive'' region at the input, but this is too
        imprecise. Our compromise solution is to use a union of two polytopes: one that includes the
        positive region $P \land
        \overrightarrow{x} \geq 0$ and another that includes as much of the non-positive region as possible. 
        Our goal then is to devise a solution that efficiently constructs a 2-polytope under-approximation 
        by only inexpensive linear algebraic manipulations, i.e., without using an LP or an SMT solver.
        We describe two methods that differ only in the way it constructs the non-positive polytope at
        the input to Relu - one that works when $P$ includes the 0 vector and another when it does not.
       
        \textbf{A note on cumulative back propagation:} If we continue to back propagate a union of two polytopes produced by Relu for each polytope across
        all layers, we will still end up with an exponential number of polytopes as we move backwards
        along the layers.
        To avoid this situation, we use the following simplification for performance.
        We keep the 2-polytope under-approximation only to perform inclusion check (line~10 in Algorithm~1).
        The polytope corresponding to the negative region is dropped before it is subsequently back propagated further into
        earlier layers.
        This simplification is sound because dropping one of the polytopes still represents an under-approximation.

        \subsubsection{$\mathit{Relu}$ Backpropagation around Zero}

        The main intuition used to construct the negative region to be included in the
        under-approximation is as follows: if $0$ is inside $P$, then the entire negative quadrant with
        all negative values will be in $\mathit{WeakestPrecond}(\mathit{Relu}, P)$, since any negative vector gets mapped to $0$ by
        $\mathit{Relu}$. Therefore, the region $ \overrightarrow{x} \leq 0$ can be included in the
        under-approximation. We try to do better by including a region of the form $\overrightarrow{x}
        \leq \overrightarrow{\eta}$, where all components of $\overrightarrow{\eta}$ are non-negative.
        A sufficient condition that such an $\overrightarrow{\eta}$ should satisfy
        is: $\forall \overrightarrow{y} 0 \leq \overrightarrow{y} \leq \overrightarrow{\eta}
        \;{\Rightarrow}\; \overrightarrow{y} L \leq \overrightarrow{u} $. We construct the most liberal
        $\overrightarrow{\eta}$ satisfying the necessary condition as per an an optimality criterion
        described below.

        The optimality measure we use is $\prod_i \eta_i$, where $\eta_i$ are the
        components of $\overrightarrow{\eta}$. Intuitively, it covers the maximum
        ``volume''. For each inequality $\overrightarrow{x}.\overrightarrow{v} \leq
        u_k$ in $P$, we can maximise this measure by finding the
        $\overrightarrow{\eta}$ at which the partial derivative of this product
        with respect to each $\eta_i$ is $0$, constrained by
        $\overrightarrow{x}.\overrightarrow{v} = u_k$.  This can be done by solving
        a set of linear equations in $\eta_i$. Thus we get one
        $\overrightarrow{\eta}$ satisfying each inequality and maximising the
        product, and taking the component wise minimum of these gives us our final
        $\overrightarrow{\eta}$. A detailed description of this process is given in
        Appendix~\ref{app:pb0_details}.
        
        This computation only involves solving a set of linear equations for each
        inequality in $P$, and then taking minimums. This can be done very
        efficiently by standard linear algebra algorithms.

        \subsubsection{$\mathit{Relu}$ Backpropagation along Quadrant with Center Point}

        When \\
        $P: \overrightarrow{x} L \leq \overrightarrow{u}$ does not contain $0$, we use the
        following method. We pick a non-positive quadrant, and add all the points in that quadrant that
        map to points in $\overrightarrow{x}L \leq \overrightarrow{u}$ via the $\mathit{Relu}$.

        Say we have a non-positive quadrant, represented by a matrix $Q$, so that $\overrightarrow{x} Q
        \leq 0$ describes the linear conditions stating that $\overrightarrow{x}$ is in the chosen quadrant.
        Now, in the given quadrant, $\mathit{Relu}$ behaves like a linear projection that sets the components
        chosen to be negative to zero. Let this projection be given by the matrix $\Pi_Q$. Then, if
        $\overrightarrow{x}$ satisfies $\overrightarrow{x} Q \leq 0$ and $\overrightarrow{x} \Pi_Q L
        \leq \overrightarrow{u}$, $\mathit{Relu}(\overrightarrow{x}) = \overrightarrow{x} \Pi_Q$ satisfies
        $\overrightarrow{x}L \leq \overrightarrow{u}$. So, it is sound to add the convex polytope
        $\overrightarrow{x} Q \leq 0 \;{\wedge}\; \overrightarrow{x} \Pi_Q L \leq \overrightarrow{u}$ to
        the under-approximation.

        The above lets us capture the negative side behavior for any given non-positive quadrant, but leaves the
        question of choosing the quadrant open.
        The choice of the quadrant to use is made based on the following  \textit{center point} heuristic: If
        the center point of $\mathit{reach[i]}$ is in a quadrant, we can have reasonably high confidence of the
        over-approximation being in that quadrant. Since we wish to find an under-approximation that has
        the best chance of containing $\mathit{reach[i]}$, we pick the quadrant which contains the center point.

        \subsection{Inclusion Checking}
        \label{subsec:incl}

            Our goal is to check whether $\mathit{reach}[i]$, given by basis $B$ and center
            $\overrightarrow{c}$, is included in $\mathit{safe}[i]$, given as union of $P_1: \overrightarrow{x} L_1
            \leq \overrightarrow{u_1}$ and $P_2: \overrightarrow{x} L_2 \leq \overrightarrow{u_2}$.

            Inclusion check is challenging because the right hand side is a disjunction of two convex polytopes.
            The fact that $\mathit{reach}[i]$ is represented as an affine region also makes the task complicated as an affine region cannot easily
            be converted into a convex polytope over $\overrightarrow{x}$. We implement inclusion check by performing
            multiple LP calls each of which is designed to be simple.
            Depending on the method used for back propagation, we have two cases. For both the cases, we
            reduce the problem to checking validity over all ($\overrightarrow{x})$of a query of the following form:

            \begin{equation*}
            \begin{aligned}
                (\exists \overrightarrow{\alpha} :\; \overrightarrow{x} = \overrightarrow{\alpha} B +
                \overrightarrow{c}
                \wedge \; | \overrightarrow{\alpha} | \leq 1
                \wedge \; \overrightarrow{x} . \overrightarrow{v} \geq k)
                \;{\Rightarrow}\; \overrightarrow{x} L \leq \overrightarrow{u}
            \end{aligned}
            \end{equation*}

            \subsubsection{Backpropagation along quadrant:}
            In this case, each of the two polytopes $P_1$and $P_2$ is entirely contained in separate quadrants. There is a hyperplane that
            separates these quadrants, let it be given by $\overrightarrow{x} . \overrightarrow{v} = k$. Each
            polytope lies entirely on one side of this hyperplane. Let's say that $\overrightarrow{x} L_1 \leq
            \overrightarrow{u_1}$ lies on the side given by $\overrightarrow{x} . \overrightarrow{v} \leq k$ and
            $\overrightarrow{x} L_2 \leq \overrightarrow{u_2}$ lies on the side given by $\overrightarrow{x} .
            (-\overrightarrow{v}) \leq k$. Then, it suffices to show that all points in $\mathit{reach}[i]$
            satisfying $\overrightarrow{x} . \overrightarrow{v} \leq k$ is in $\overrightarrow{x} L_1 \leq
            \overrightarrow{u_1}$, and all points satisfying $\overrightarrow{x} .( -\overrightarrow{v}) \leq k$
            is in $\overrightarrow{x} L_2 \leq \overrightarrow{u_2}$. This gives us two queries of the above form.

            \subsubsection{Backpropagation around $0$:}
            In this case if $\overrightarrow{x}$ does not satisfy any constraint in $\overrightarrow{x} \leq
            \overrightarrow{\eta}$, it must be in the positive side polytope $\overrightarrow{x} L_1 \leq \overrightarrow{u_1}$.
            Thus, for each constraint in $\overrightarrow{x} \leq \overrightarrow{\eta}$, we check if all $\overrightarrow{x}$ in
            $\mathit{reach}[i]$ that do not satisfy the constraint are included the convex polytope. This gives
            us $n$ checks of the above form for $n$ dimensional $\overrightarrow{x}$.

            \subsubsection{Solving the above query:}
            To solve the earlier validity query, we negate it and then substitute
            $\overrightarrow{x}$ from $\overrightarrow{x} = \overrightarrow{\alpha} B + \overrightarrow{c}$ to
            reduce the other constraints to one on $\overrightarrow{\alpha}$ to get the following UNSAT query:

            \begin{equation*}
            \begin{aligned}
                | \overrightarrow{\alpha} | \leq 1
                \; \wedge \; \overrightarrow{\alpha} B . \overrightarrow{v} \geq k - \overrightarrow{c} . \overrightarrow{v}
                \;\wedge\; \neg ( \overrightarrow{\alpha} B L \leq \overrightarrow{u} - \overrightarrow{c}
                L)
            \end{aligned}
            \end{equation*}

            This check can be done by taking each constraint in the convex polytope $\overrightarrow{\alpha}
            B L \leq \overrightarrow{u} - \overrightarrow{c} L$ and checking if an $\overrightarrow{\alpha}$
            within $| \overrightarrow{\alpha} | \leq 1 \; \wedge \; \overrightarrow{\alpha} B$ violates it. Each
            such query is passed to an LP solver. This means that in the worst case there are as many LP calls
            as $n$ times the number of basis vectors in $B$. However, each LP call only has one linear
            constraint, the other constraints simply bound the value of $\overrightarrow{\alpha}$, and thus can
            be solved very efficiently.

        \subsection{Counterexample Checking}
        \label{subsec:spur}

            If the inclusion fails, we obtain a point that witnesses the violation of the inclusion (of the
            over-approximation in the under-approximation, at some layer). We wish to find several
            approximate pullbacks \footnote{When we use the term ``pullback'' in this section, it refers to the
                pullback of the \emph{counterexample} to get a finite list of points, and should not be
                confused with pullback of $\mathit{safe}[i]$} of this layer by layer until the input layer,
            giving us several potential counterexamples. This is what $\mathit{pullBackCex}$ does, given a
            point $\overrightarrow{z_{j+1}}$ at layer $j+1$ and an over-approximation $A_j$ at layer $j$, it
            returns a list of points in $A_j$ which lead to points closer than $D$ (in terms of euclidean
            distance) to $\overrightarrow{z_{j+1}}$ under the action of layer $j$. Then, we can use this
            function repeatedly layer by layer to obtain the set of potential counterexamples at the input.
            Here, $D$ is a parameter that we tune.

            Note that the pullback of $\overrightarrow{z_{j+1}}$ across the linear layer given by $W_j$ and
            $\overrightarrow{b_j}$ is the $\overrightarrow{x}$ satisfying $\overrightarrow{x} W_j +
            \overrightarrow{b_j} - \overrightarrow{z_{j+1}} = 0$. Pulling this region back over
            $\mathit{Relu}$ is in general hard, since there are potentially exponentially many quadrants to
            consider. However, if we pick a quadrant, the action of $\mathit{Relu}$ reduces to a linear
            transform setting the negative axes to $0$. Then, if this linear transform is $\Pi$, the
            pullback within this quadrant are the $\overrightarrow{x}$ satisfying $\overrightarrow{x} \Pi W_j +
            \overrightarrow{b_j} - \overrightarrow{z_{j+1}} = 0$.

            Now, say $A_j$ is given by $\exists \overrightarrow{\alpha} :\; \overrightarrow{x} =
            \overrightarrow{\alpha} B_j + \overrightarrow{c_j}, \; | \overrightarrow{\alpha} | \leq
            \overrightarrow{b_j}$. Then, we generate several \footnote{around $10000$ in our current implementation}
            possible values of $|\overrightarrow{\alpha}| \leq 1$ randomly. Corresponding to these we get several
            $\overrightarrow{z_j} = \overrightarrow{\alpha} B_j + \overrightarrow{c_j}$ in $A_j$. For each
            such $\overrightarrow{z_j}$, we find the pullback of $\overrightarrow{z_{j+1}}$ in the quadrant
            to which $\overrightarrow{z_j}$ belongs using the method described above. This gives us a large
            list of points in $A_j$ from a variety of different quadrants. 
            
            The probability of any of the $\overrightarrow{z_j}$ generated being an approximate pullback is
            still quite low, as they have essentially been randomly generated. To improve this probability,
            we move the  $\overrightarrow{\alpha}$ in the direction which causes $\overrightarrow{\alpha}
            B_j \Pi W_j + \overrightarrow{c_j} \Pi W_j + \overrightarrow{b_j} - \overrightarrow{z_{j+1}} =
            \overrightarrow{z_j} \Pi W_j + \overrightarrow{b_j} - \overrightarrow{z_{j+1}}$ to come closest
            to $0$ as far as we can without the bounds on $\alpha$ being violated. This brings the
            $\mathit{Relu}$ image of $\overrightarrow{z_j}$ as close to $\overrightarrow{z_{j+1}}$ as
            possible. We now simulate the $\overrightarrow{z_j}$ for one layer, and discard those that lead
            to points farther than $D$ from $\overrightarrow{z_{j+1}}$ under the action of the layer, to get
            our required list of points. Note that this step can end up eliminating all the
            $\overrightarrow{z_j}$, in which case the approximate pullback of the counterexample fails.

            As we repeat this process for previous layers, the distance of the approximate pullbacks from
            the original counterexample increases by a factor of $D$. To counteract this, we scale $D$ down
            by the number of layers at the beginning of the algorithm.

        \subsection{Example (continued from Sect.~\ref{subsec:example-preprocess})}
        \label{subsec:example-propagate}

            \begin{figure}
                \centering
                \begin{tikzpicture}[
                plain/.style={
                    draw=none,
                    fill=none,
                    },
                net/.style={
                    nodes={
                        draw,
                        circle,
                        inner sep=1pt
                        },
                    column sep=0.4cm,
                    row sep=0.2cm
                    },
                pnt/.style={
                    fill,
                    circle,
                    inner sep=0pt,
                    outer sep=0pt,
                    minimum size=3pt
                },
                >=latex
                ]
                \matrix[net] (mat)
                {
                    |[plain]| &[1.2cm] 
                        \node (p10) [pnt] {}; & 
                        \node (y00) [dashed, label={[yshift=-0.1cm] $\mathit{500}$}] {$y_{00}$}; &[0.5cm] 
                        \node [label={[yshift=-0.1cm] $\mathit{500}$}] (x10) {$x_{10}$}; &[1.2cm] 
                        |[plain]| & |[plain]| &[0.5cm] |[plain]| \\
                    \node (x00) [label={[yshift=-0.1cm] $\mathit{0.5}$}] {$x_{00}$}; &[1.2cm] 
                        \node (p11) [pnt] {}; & 
                        \node (y01) [dashed, label={[yshift=-0.3cm] $\mathit{-500}$}] {$y_{01}$}; &[0.5cm] 
                        \node (x11) [label=$\mathit{0}$] {$x_{11}$}; &[1.2cm] \node (p30) [pnt] {}; & 
                        \node (y10) [dashed, label=$\mathit{1}$] {$y_{10}$}; &[0.5cm] 
                        \node (x20) [label=$\mathit{1}$] {$x_{20}$};\\
                    \node (x01) [label=$\mathit{0}$] {$x_{01}$}; &[1.2cm] 
                        \node (p12) [pnt] {}; & 
                        \node (y02) [dashed, label={[yshift=-0.2cm] $\mathit{499}$}] {$y_{02}$}; &[0.5cm] 
                        \node (x12) [label={[yshift=-0.2cm] $\mathit{499}$}] {$x_{12}$}; &[1.2cm] 
                        \node (p31) [pnt] {}; & \node (y11) [dashed, label=$\mathit{0}$] {$y_{11}$}; &[0.5cm] 
                        \node (x21) [label=$\mathit{0}$] {$x_{21}$};\\
                    |[plain]| &[1.2cm] \node (p13) [pnt] {}; & 
                        \node (y03) [dashed, label={[yshift=-0.3cm] $\mathit{-501}$}] {$y_{03}$}; &[0.5cm] 
                        \node (x13) [label=$\mathit{0}$] {$x_{13}$}; &[1.2cm] 
                        |[plain]| & |[plain]| &[0.5cm] |[plain]| \\
                };
                \draw[->] (x00) -- (p10) node[pos=0.8, above] {$g$};
                \draw[->] (x00) -- (p11) node[pos=0.8, above] {$-g$};
                \draw[->] (x00) -- (p12) node[pos=0.8, above] {$g$};
                \draw[->] (x00) -- (p13) node[pos=0.8, above] {$-g$};
                \draw[->] (x01) -- (p10) node[pos=0.8, below] {$-g$};
                \draw[->] (x01) -- (p11) node[pos=0.8, below] {$g$};
                \draw[->] (x01) -- (p12) node[pos=0.8, below] {$-g$};
                \draw[->] (x01) -- (p13) node[pos=0.8, below] {$g$};
                \draw[->] (p10) -- (y00) node[pos=0.5, sloped, below] {$0$};
                \draw[->] (p11) -- (y01) node[pos=0.5, sloped, below] {$0$};
                \draw[->] (p12) -- (y02) node[pos=0.5, sloped, below] {$-1$};
                \draw[->] (p13) -- (y03) node[pos=0.5, sloped, below] {$-1$};
                \draw[->] (y00) -- (x10) node[pos=0.5, sloped, below] {$\mathit{Relu}$};
                \draw[->] (y01) -- (x11) node[pos=0.5, sloped, below] {$\mathit{Relu}$};
                \draw[->] (y02) -- (x12) node[pos=0.5, sloped, below] {$\mathit{Relu}$};
                \draw[->] (y03) -- (x13) node[pos=0.5, sloped, below] {$\mathit{Relu}$};
                \draw[->] (x10) -- (p30) node[near end, below] {$1$};
                \draw[->] (x12) -- (p30) node[near end, below] {$-1$};
                \draw[->] (x11) -- (p31) node[near end, below] {$1$};
                \draw[->] (x13) -- (p31) node[near end, below] {$-1$};
                \draw[->] (p30) -- (y10) node[pos=0.5, sloped, below] {$0$};
                \draw[->] (p31) -- (y11) node[pos=0.5, sloped, below] {$0$};
                \draw[->] (y10) -- (x20) node[pos=0.5, sloped, below] {$\mathit{Relu}$};
                \draw[->] (y11) -- (x21) node[pos=0.5, sloped, below] {$\mathit{Relu}$};
                \end{tikzpicture}
                \caption{$\sigma = (0 {\rightarrow} 1, 1 {\rightarrow} 0)$, $g=1000$}
                \label{eg1}
            \end{figure}

            Consider the neural network shown in Fig.\ref{eg1}. Here, we have separated the result of
            computing the weighted sum from that of the application of the $\mathit{Relu}$ into separate
            nodes, represented by dashed and solid circles respectively. Also, we show the weights as labels
            on the arrows coming into a combination point, and biases as labels of arrows emerging from the
            point.  The arrows for weights that are $0$ have been omitted. The values at (output of) each node in the
            network for the input in the range $[0.5 \; 0]$ are shown in the diagram at that node.
            
            This network has the following symmetry property: $0 \leq x_{00}, x_{01}, x_{00}',
            x_{01}' \leq 1 \;{\wedge}\; x_{00} = x_{01}' \;{\wedge}\; x_{01} = x_{00}' \;{\Rightarrow}\;
            |[x_{40} \; x_{41}] - [x_{40}' \; x_{41}']| \leq 0.1$. This expresses the fact that flipping the
            inputs leads to the outputs being flipped.

            \subsubsection{Preprocessing:}
            
            The $W_i$ and $\overrightarrow{b_i}$ are calculated as follows: If the weights and bias
            of ayer $i$ are $W^o$ and $\overrightarrow{b^o}$, then $W_i = \bigl[\begin{smallmatrix} W^0 & 0 \\ 0 & W^0
            \end{smallmatrix}\bigr]$ and $\overrightarrow{b_i} = [\overrightarrow{b^o} ~
            \overrightarrow{b^o}]$ as we need to track both the original and permuted values at each layer. For this example we have:
            
            \begin{equation*}
            \begin{matrix}
                \begin{aligned}
                    &W_0 = \\
                    &\begin{bmatrix} 
                        1000 & -1000 & 1000 & -1000 & 0 & 0 & 0 & 0 \\
                        -1000 & 1000 & -1000 & 1000 & 0 & 0 & 0 & 0 \\
                        0 & 0 & 0 & 0 & 1000 & -1000 & 1000 & -1000 \\
                        0 & 0 & 0 & 0 & -1000 & 1000 & -1000 & 1000 \\
                    \end{bmatrix} \\
                    &\overrightarrow{b_0} = \begin{bmatrix} 0 & 0 & -1 & -1 & 0 & 0 & -1 & -1 \\
                    \end{bmatrix} \\
                \end{aligned} &
                \begin{aligned}
                    W_1 &= \begin{bmatrix} 
                         1 &  0 &  0 &  0 \\
                         0 &  1 &  0 &  0 \\
                        -1 &  0 &  0 &  0 \\
                         0 & -1 &  0 &  0 \\
                         0 &  0 &  1 &  0 \\
                         0 &  0 &  0 &  1 \\
                         0 &  0 & -1 &  0 \\
                         0 &  0 &  0 & -1 \\
                    \end{bmatrix} \\ 
                    \overrightarrow{b_1} &= \begin{bmatrix} 0 & 0 & 0 & 0 \\ \end{bmatrix} \\
                \end{aligned}
            \end{matrix}
            \end{equation*}

            \subsubsection{Action of $\mathit{initPre}$ and $\mathit{initPost}$:}

            Now, $\mathit{initPre}$ calculates $\mathit{reach}[0]$ as the following affine region given by basis
            $B_0$ and center $\overrightarrow{c_0}$, and $\mathit{initPost}$ expresses $\mathit{safe}[2]$ as
            a convex polytope:
            
            \begin{equation}
            \begin{matrix}
                \begin{aligned}
                    &\mathbf{reach[0]}: \\
                    \exists \overrightarrow{\alpha} : [\overrightarrow{x_0} ~ \overrightarrow{x_0'}] &= \overrightarrow{\alpha} B_0 + \overrightarrow{c_0}, \;
                    |\overrightarrow{\alpha}| \leq 1 \\ 
                    B_0 &= \begin{bmatrix} 0.5 & 0 & 0 & 0.5 \\ 0 & 0.5 & 0.5 & 0 \\ \end{bmatrix}\\
                    \overrightarrow{c_0} &= \begin{bmatrix} 0.5 & 0.5 & 0.5 & 0.5 \\ \end{bmatrix}
                \end{aligned} &
                \begin{aligned}
                    &\mathbf{safe[2]}: \\
                    [\overrightarrow{x_4} ~ \overrightarrow{x_4'}] \begin{bmatrix} 
                        1 & 0 & -1 & 0 \\ 
                        0 & 1 & 0 & -1 \\ 
                        0 & -1 & 0 & 1 \\ 
                        -1 & 0 & 1 & 0 \\ 
                    \end{bmatrix} &\leq 
                    \begin{bmatrix} 0.1 & 0.1 & 0.1 & 0.1 \end{bmatrix}
                \end{aligned}
                \label{eqn:prepost} \\
            \end{matrix}
            \end{equation}

            \subsubsection{Forward Propagation:} 
         
            $\mathit{ForwardPropagate}$ then propagates (\ref{eqn:prepost}) across the layers to get affine regions
            that are over-approximations for the reachable region for that layer. While propagation across
            the linear layer can be done easily via matrix multiplication, propagating across the $\mathit{Relu}$ layer is in general is hard,
            since we need to take into account all possible branching behaviors. We do this via a tie class
            analysis (section~\ref{subsec:forprop}) that exploits the inherent symmetry of the network and precondition. For this network,
            propagating across the first linear layer gives us an affine region given by the basis and
            center:

            \begin{equation*}
            \begin{aligned}
                B_0' &= \begin{bmatrix}  
                    500 & -500 &  500 & -500 & -500 &  500 & -500 &  500 \\ 
                   -500 &  500 & -500 &  500 &  500 & -500 &  500 & -500 \\ 
                \end{bmatrix} \\
                \overrightarrow{c_0'} &= \begin{bmatrix} 0 & 0 & -1 & -1 & 0 & 0 & -1 & -1 \\ \end{bmatrix} \\
            \end{aligned}
            \end{equation*}

            Then, propagating across the $\mathit{Relu}$ using the tie class analysis (section~\ref{subsec:forprop}) gives us the basis
            $B_1$  and center $\overrightarrow{c_1}$ for $\mathit{reach[1]}$. Similarly, the algorithm
            propagates across the second layer to get $B_1'$, $\overrightarrow{c_1'}$, $B_2$ and
            $\overrightarrow{c_2}$. In this case, the affine region before and after the $\mathit{Relu}$
            turn out to be the same, and there is no loss in precision going from $B_1'$ to $B_2$. The matrices are:

            \begin{equation}
            \label{eqn:eg_mats}
            \begin{aligned}
                & B_1 = &\;& B_1', B_2 = \\
                & \begin{bmatrix}  
                    500 &    0 &    0 &    0 &    0 &  500 &    0 &    0 \\ 
                   -500 &    0 &    0 &    0 &    0 & -500 &    0 &    0 \\ 
                      0 & -500 &    0 &    0 & -500 &    0 &    0 &    0 \\ 
                      0 &  500 &    0 &    0 &  500 &    0 &    0 &    0 \\ 
                      0 &    0 &  500 &    0 &    0 &    0 &    0 &  500 \\ 
                      0 &    0 & -500 &    0 &    0 &    0 &    0 & -500 \\ 
                      0 &    0 &    0 & -500 &    0 &    0 & -500 &    0 \\ 
                      0 &    0 &    0 &  500 &    0 &    0 &  500 &    0 \\ 
                \end{bmatrix} &\;&
                \begin{bmatrix} 
                     500 &     0 &     0 &  -500 \\ 
                    -500 &     0 &     0 &   500 \\ 
                       0 &  -500 &  -500 &     0 \\ 
                       0 &   500 &   500 &     0 \\ 
                    -500 &     0 &     0 &  -500 \\ 
                     500 &     0 &     0 &   500 \\ 
                       0 &   500 &   500 &     0 \\ 
                       0 &  -500 &  -500 &     0 \\ 
                \end{bmatrix}\\
                & \overrightarrow{c_1} = \begin{bmatrix} 0 & 0 & 0 & 0 & 0 & 0 & 0 & 0 \\ \end{bmatrix} &\;&
                \overrightarrow{c_1'}, \overrightarrow{c_2} = \begin{bmatrix} 0 & 0 & 0 & 0 \\ \end{bmatrix} \\
            \end{aligned}
            \end{equation}

            \subsubsection{Inclusion check:}
            
            Now, we see that if we substitute $\overrightarrow{x}$ with the form given in $\mathit{reach}[2]$ into
            $\mathit{safe}[2]$ (which is the postcondition $|[x_{40} \; x_{41}] - [x_{40}' \; x_{41}']| \leq 0.1$),
            the right side of the inequality simplifies to $0$. So, $\mathit{reach}[2]$ is included in
            $\mathit{safe}[2]$.  This is done by an algorithm (section~\ref{subsec:incl}) that checks this using an LP  solver, and since
            it succeeds in this case, it returns \textit{property holds}. 
            
            Note that for this example, it was unnecessary to perform any back propagation of the
            $\mathit{safe}[i]$ to previous layers, as the inclusion check succeeded at the output layer.
            In general, back propagation (section~\ref{subsec:backprop}) would be performed to compute under-approximations.
            Spuriousness check (section~\ref{subsec:spur}) will be needed if the inclusion check fails.

\section{Experiments}
\label{sec:experiment}

    We have demonstrated our algorithm using a prototype implementation written
    in python\footnote{We will be submitting an artifact containing all the
    benchmarks, our implementation, and the scripts to reproduce our
    experimental results.}. As an additional step after tie-class analysis, we
    optimise the basis obtained by removing linearly dependent vectors using
    singular value decomposition. Numerical calculations including matrix
    multiplication and singular value decomposition were done using the
    \textit{numpy} library. For the inclusion checks, the LP solver provided in
    the \textit{scipy} library was used. All times are in reported here are in
    seconds. All the experiments were run on an Intel i7 9750H processor with 6
    cores and 12 threads with 32 GB RAM.

    We have compared our algorithm with the
    Marabou~\cite{DBLP:conf/cav/KatzHIJLLSTWZDK19,DBLP:conf/sigcomm/KazakBKS19}
    implementation of the Reluplex~\cite{DBLP:conf/cav/KatzBDJK17} on a few
    DNNs of various sizes with the following target behavior: for $n$ inputs,
    there should be $n$ outputs so that if input $i$ is the largest among all
    the inputs, output $i$ should be $1$. These networks have three layers
    excluding the input layer, with sizes $2n(n-1)$, $n(n-1)$ and $n$
    respectively. We check the following permutation invariance property:
    \begin{equation*}
    \begin{aligned}
    0 \leq \overrightarrow{x} \leq 1 \;{\Rightarrow}\; |
    {\sigma}(\mathcal{N}(\overrightarrow{x})) - \mathcal{N}({\sigma}(\overrightarrow{x}))
    | \leq \epsilon
    \end{aligned}
    \end{equation*}
    
    Where $\sigma$ represents the permutation sending $1 \;{\rightarrow}\; 2, 2 \;{\rightarrow}\; 3
    {\cdots} n \;{\rightarrow}\; 1$ cyclically, and $\epsilon$ varies across the experiments. Note
    that if the network follows the target behavior exactly, then this property should hold.
    
    We first demonstrate our algorithm on a set of hand-crafted networks for
    which we have manually fixed the weights. The first layer has two neurons
    $p_{i,j}$ and $q_{i,j}$ for each pair $(i, j)$ of inputs. The input to the
    $\mathit{Relu}$ for these neurons are $1000(i-j)$ and $1000(i-j)-1$
    respectively. The second layer has one neuron $r_{i,j}$ for each pair $(i,
    j)$ of inputs, and the input to it's $\mathit{Relu}$ is $p_{i,j} -
    q_{i,j}$. The output layer has one output $s_i$ for each input $i$, with
    $2\sum_j r_{i,j} - 2n + 3$ being fed into it's $\mathit{Relu}$.
    Intuitively, $r_{i,j}$'s output is designed to be $1$ whenever $i$ is
    considerably bigger than $j$, and $0$ otherwise, and $s_i$ calculates
    $\wedge_j (i > j)$, achieving the desired behavior.

    This hand-crafted DNN for this example is quite symmetric, in that the intermediate calculations
    being performed are symmetrically linked to the input. That is, permuting
    the inputs leads to a more complicated permutation of the intermediate
    layers. Thus, intuitively it should be easy to to prove the symmetry
    property we have described. However, as the input to the network varies
    within the precondition region, the input to the $\mathit{Relu}$ nodes
    regularly switch between positive and negative. This can potentially lead
    to exponentially many case-splits unless an effective abstraction is used.
    Table~\ref{tab:compsafe} compares the time taken by our algorithm and by
    Marabou on these networks. Our algorithm converges very quickly in every
    case whereas Marabou times out (after 100 seconds) in all but the smallest
    case. The test result demonstrates that the over-approximation and under-approximation
    used in our algorithm form an effective abstraction for this example, and is
    likely to be so for similar, symmetric networks. Note that in the table
    when a time is given as $>t$, it is denoting that a timeout was declared
    after $t$ seconds.

   \begin{table}
	   \centering
    \caption{Comparison of Marabou and Our Algorithm on Safe Synthetic Networks}
	   \label{tab:compsafe}
    \setlength{\tabcolsep}{8pt}
    \begin{tabular}{ | c | c | c | c | c | }
        \hline
        Number & Size & Our &
        \multicolumn{2}{|c|}{ Marabou } \\
        of Inputs & of Network & Algorithm & Time & Splits \\
        \hline
        $3  $&$ 21  $&$ 0.074 $&$  4.833 $&$  2046  $\\
        $4  $&$ 40  $&$ 0.112 $&$ >100.8 $&$ >11234 $\\
        $5  $&$ 65  $&$ 0.163 $&$ >101.9 $&$ >5186  $\\
        $6  $&$ 96  $&$ 0.269 $&$ >100.1 $&$ >2243  $\\
        $7  $&$ 133 $&$ 0.493 $&$ >106.8 $&$ >1533  $\\
        $8  $&$ 176 $&$ 0.911 $&$ >126.5 $&$ >475   $\\
        $9  $&$ 225 $&$ 1.477 $&$ >183.9 $&$ >467   $\\
        $10 $&$ 280 $&$ 2.276 $&$ >158.7 $&$ >394   $\\
        \hline
    \end{tabular} 
   \end{table}

    We also test our algorithm on an unsafe problem using the same hand-crafted
    network from the previous example. To do so, we change the permutation on
    the output side to be the identity permutation. As permuting the inputs
    cyclically should not leave the outputs unchanged, we should always find a
    counterexample in these tests. The results are given in
    Table~\ref{tab:compunsafe} and show that our counterexample search is able
    to find counterexamples in a way that is competitive with Marabou,
    especially for networks with $8$ or more inputs.
    
   \begin{table}
	   \centering
    \caption{Comparison of Marabou and Our Algorithm on Unsafe Synthetic Networks}
	   \label{tab:compunsafe}
    \setlength{\tabcolsep}{8pt}
    \begin{tabular}{ | c | c | c | c | c | }
        \hline
        Number & Size & Our &
        \multicolumn{2}{|c|}{ Marabou } \\
         of Inputs & of Network & Algorithm & Time & Splits \\
        \hline
        $3  $&$ 21  $&$ 0.048 $&$ 0.187 $&$ 68  $\\
        $4  $&$ 40  $&$ 0.074 $&$ 0.202 $&$ 38  $\\
        $5  $&$ 65  $&$ 0.132 $&$ 0.267 $&$ 47  $\\
        $6  $&$ 96  $&$ 0.233 $&$ 0.603 $&$ 60  $\\
        $7  $&$ 133 $&$ 0.422 $&$ 1.085 $&$ 64  $\\
        $8  $&$ 176 $&$ 0.809 $&$ 71.89 $&$ 299 $\\
        $9  $&$ 225 $&$ 1.508 $&$ 5.011 $&$ 91  $\\
        $10 $&$ 280 $&$ 2.157 $&$ 29.09 $&$ 202 $\\
        \hline
    \end{tabular}
   \end{table}

   \begin{table}
	   \centering
	   \caption{Comparison of Marabou and Our Algorithm on Trained Networks (INCONS denotes that the algorithm returned \emph{inconclusive}; TO denotes a \emph{timeout}; SAFE denotes that the property was proved; CEX denotes that the property was refuted and a counterexample was returned)}
	   \label{tab:comptrained}
    \setlength{\tabcolsep}{8pt}
    \begin{tabular}{ | c | c | c | c | c | c | c | c | c | }
        \hline
        \multicolumn{4}{|c|}{Network} & \multicolumn{2}{|c|}{ Our Algorithm } &
        \multicolumn{3}{|c|}{ Marabou } \\
        $n$ & Size & $\epsilon$ & Accuracy & Time & Result & Time & Splits & Result \\
        \hline
        $ 3  $&$ 21  $&$ 0.1 $&$ 94.0\% $&$ 0.023 $& CEX    &$ 0.023  $&$ 10   $& CEX \\
        $ 3  $&$ 21  $&$ 0.3 $&$ 99.5\% $&$ 0.109 $& CEX    &$ 0.028  $&$ 15   $& CEX \\
        $ 3  $&$ 21  $&$ 0.5 $&$ 100\%  $&$ 0.249 $& INCONS &$ 0.034  $&$ 16   $& CEX \\
        $ 3  $&$ 21  $&$ 0.7 $&$ 100\%  $&$ 0.158 $& INCONS &$ 1.335  $&$ 320  $& SAFE \\
        $ 3  $&$ 21  $&$ 0.9 $&$ 100\%  $&$ 0.204 $& INCONS &$ 1.330  $&$ 274  $& SAFE \\
        $ 4  $&$ 40  $&$ 0.1 $&$ 98.3\% $&$ 0.057 $& CEX    &$ 0.375  $&$ 47   $& CEX \\
        $ 4  $&$ 40  $&$ 0.3 $&$ 99.1\% $&$ 0.114 $& CEX    &$ 0.495  $&$ 51   $& CEX \\
        $ 4  $&$ 40  $&$ 0.5 $&$ 99.8\% $&$ 0.112 $& CEX    &$ 0.464  $&$ 50   $& CEX \\
        $ 4  $&$ 40  $&$ 0.7 $&$ 99.9\% $&$ 0.110 $& CEX    &$ 0.426  $&$ 48   $& CEX \\
        $ 4  $&$ 40  $&$ 0.9 $&$ 100\%  $&$ 0.119 $& CEX    &$ 0.418  $&$ 49   $& CEX \\
        $ 5  $&$ 65  $&$ 0.1 $&$ 97.1\% $&$ 0.197 $& CEX    &$ 0.684  $&$ 35   $& CEX \\
        $ 5  $&$ 65  $&$ 0.3 $&$ 98.5\% $&$ 0.209 $& CEX    &$ 0.708  $&$ 35   $& CEX \\
        $ 5  $&$ 65  $&$ 0.5 $&$ 99.5\% $&$ 0.188 $& CEX    &$ 0.682  $&$ 35   $& CEX \\
        $ 5  $&$ 65  $&$ 0.7 $&$ 99.7\% $&$ 0.202 $& CEX    &$ 0.699  $&$ 36   $& CEX \\
        $ 5  $&$ 65  $&$ 0.9 $&$ 99.9\% $&$ 0.224 $& CEX    &$ 21.26  $&$ 563  $& CEX \\
        $ 6  $&$ 96  $&$ 0.1 $&$ 98.0\% $&$ 0.012 $& CEX    &$ 3.070  $&$ 85   $& CEX \\
        $ 6  $&$ 96  $&$ 0.3 $&$ 98.6\% $&$ 0.018 $& CEX    &$ 3.138  $&$ 85   $& CEX \\
        $ 6  $&$ 96  $&$ 0.5 $&$ 99.2\% $&$ 0.014 $& CEX    &$ 3.314  $&$ 85   $& CEX \\
        $ 6  $&$ 96  $&$ 0.7 $&$ 99.4\% $&$ 0.013 $& CEX    &$ 3.327  $&$ 85   $& CEX \\
        $ 6  $&$ 96  $&$ 0.9 $&$ 99.7\% $&$ 0.012 $& CEX    &$ 3.292  $&$ 85   $& CEX \\
        $ 7  $&$ 133 $&$ 0.1 $&$ 87.5\% $&$ 0.011 $& CEX    &$ 5.651  $&$ 84   $& CEX \\
        $ 7  $&$ 133 $&$ 0.3 $&$ 96.1\% $&$ 0.012 $& CEX    &$ 5.810  $&$ 86   $& CEX \\
        $ 7  $&$ 133 $&$ 0.5 $&$ 98.1\% $&$ 1.153 $& CEX    &$ 5.991  $&$ 84   $& CEX \\
        $ 7  $&$ 133 $&$ 0.7 $&$ 98.7\% $&$ 1.081 $& CEX    &$ 5.957  $&$ 84   $& CEX \\
        $ 7  $&$ 133 $&$ 0.9 $&$ 99.5\% $&$ 1.229 $& CEX    &$ 113.6  $&$ 531  $& CEX \\
        $ 8  $&$ 176 $&$ 0.1 $&$ 65.7\% $&$ 0.012 $& CEX    &$ 44.42  $&$ 258  $& CEX \\
        $ 8  $&$ 176 $&$ 0.3 $&$ 68.5\% $&$ 1.584 $& CEX    &$ 42.80  $&$ 258  $& CEX \\
        $ 8  $&$ 176 $&$ 0.5 $&$ 71.3\% $&$ 1.586 $& CEX    &$ 43.60  $&$ 258  $& CEX \\
        $ 8  $&$ 176 $&$ 0.7 $&$ 73.1\% $&$ 1.630 $& CEX    &$ 43.23  $&$ 258  $& CEX \\
        $ 8  $&$ 176 $&$ 0.9 $&$ 75.6\% $&$ 1.550 $& CEX    &$ 44.60  $&$ 258  $& CEX \\
        $ 9  $&$ 255 $&$ 0.1 $&$ 58.4\% $&$ 1.193 $& CEX    &$ >120.3 $&$ >228 $& TO  \\
        $ 9  $&$ 255 $&$ 0.3 $&$ 70.2\% $&$ 1.310 $& CEX    &$ >127.9 $&$ >179 $& TO  \\
        $ 9  $&$ 255 $&$ 0.5 $&$ 83.3\% $&$ 1.336 $& CEX    &$ >132.5 $&$ >225 $& TO  \\
        $ 9  $&$ 255 $&$ 0.7 $&$ 93.3\% $&$ 1.350 $& CEX    &$ >131.6 $&$ >215 $& TO  \\
        $ 9  $&$ 255 $&$ 0.9 $&$ 98.4\% $&$ 2.159 $& CEX    &$ >133.5 $&$ >217 $& TO  \\
        $ 10 $&$ 280 $&$ 0.1 $&$ 20.8\% $&$ 4.040 $& CEX    &$ >130.4 $&$ >58  $& TO  \\
        $ 10 $&$ 280 $&$ 0.3 $&$ 31.0\% $&$ 3.966 $& CEX    &$ >125.0 $&$ >58  $& TO  \\
        $ 10 $&$ 280 $&$ 0.5 $&$ 39.7\% $&$ 4.100 $& CEX    &$ >124.3 $&$ >58  $& TO  \\
        $ 10 $&$ 280 $&$ 0.5 $&$ 50.5\% $&$ 3.991 $& CEX    &$ >130.1 $&$ >58  $& TO  \\
        $ 10 $&$ 280 $&$ 0.5 $&$ 62.3\% $&$ 4.063 $& CEX    &$ >125.4 $&$ >58  $& TO  \\
        \hline
    \end{tabular}
   \end{table}

    Finally, we compare the performance of our algorithm with Marabou on some
    DNNs for the same problem that have been trained using stochastic gradient descent to have the
    target behavior described above. A large number of randomly generated input
    points and the corresponding $1$-hot output vectors were used as the
    training data. Note that unlike the hand-crafted networks, for the trained
    networks we have no guarantee that the property will hold. The accuracy
    column tracks the percentage of test inputs for which the property holds.
    We compare the algorithms on trained networks of various sizes, and with
    various values of $\epsilon$. The results (Table~\ref{tab:comptrained})
    show that our algorithm compares quite favorably with Marabou, especially
    as the network size increases. Though small in number, our benchmarks are
    challenging due to their size and complexity of verification.  We attribute
    the efficiency of our approach to a number of design elements that are
    crucial in our approach -- a layer by layer analysis, abstractions (that
    help reduce case-splits), under-approximations (that lead to good
    counterexamples), algebraic manipulations instead of LP/SMT calls, etc.  A
    downside of our algorithm is that it may sometimes return
    \emph{inconclusive}. A counterexample-guided refinement procedure can help
    tackle this issue.

\section{Related Work}
\label{sec:related}

    The field of DNN verification has gained significant attention in the last
    several years. DNNs are being used more and more in safety- and
    business-critical systems, and therefore it becomes crucial to formally argue
    that the presence of ML components do not compromise on the essential and
    desirable system-properties. Efforts in formal verification of neural networks
    have relied on
    abstraction-refinement~\cite{abs-refine-early-cav10,dnnAbs-CAV2020,DBLP:conf/nips/PrabhakarA19},
    constraint-solving~\cite{DBLP:conf/atva/Ehlers17,Tjeng2019EvaluatingRO,DBLP:conf/nfm/DuttaJST18,DBLP:conf/kr/AkitundeLMP18},
    abstract
    interpretation~\cite{absInt-ai2,DBLP:conf/nips/SinghGMPV18,DBLP:journals/pacmpl/SinghGPV19},
    layer-by-layer
    search~\cite{DBLP:conf/cav/HuangKWW17,DBLP:conf/tacas/WickerHK18}, two-player
    games~\cite{DBLP:journals/tcs/WuWRHK20}, and several other
    approaches~\cite{DBLP:conf/atva/JacobyBK20,DBLP:journals/tnn/XiangTJ18}.

    The most closely related work to ours is using a DNN verification engine such
    as Reluplex~\cite{DBLP:conf/cav/KatzBDJK17} and
    Marabou~\cite{DBLP:conf/cav/KatzHIJLLSTWZDK19,DBLP:conf/sigcomm/KazakBKS19} to
    verify permutation invariance properties by reasoning over two copies of the
    network.  Reasoning over multiple copies also comes up in the context of
    verifying Deep Reinforcement Learning
    Systems~\cite{DBLP:conf/sigcomm/EliyahuKKS21}.
    However, verification of DNNs is
    worst-case exponential in the size of the network and therefore our proposal to handle permutation
    invariance directly (instead of multiplying the network-size) holds a lot of
    promise.

    Polytope propagation has been quite useful in the context of neural network
    verification (e.g~\cite{syrenn,DBLP:conf/ecai/ZhangSGGLN20}). In the case of
    forward propagation, however, it requires computing the convex hull each time,
    which is an expensive operation. In contrast, our tie-class analysis helps us
    propagate the affine regions efficiently. In the backward direction, even
    though we rely on convex polytope propagation, we mitigate the worst-case
    exponential blow-up by using a 2-polytope under-approximation method that does
    not depend on LP or SMT solving, and is both scalable and effective.

    In general, the complexity of a verification exercise can be mitigated by
    abstraction-refinement techniques; in particular, technique such
    as~\cite{dnnAbs-CAV2020} for DNN verification. The essential idea is to let go
    of an exact computation, which is achieved by merging of neurons
    in~\cite{dnnAbs-CAV2020}. In~\cite{DBLP:conf/nips/PrabhakarA19}, the authors
    propose construction of a simpler neural network with fewer neurons, using
    interval weights such that the simplified network over-approximates the output
    range of the original neural network. Our work is similar in spirit, in that it
    avoids exact computation unless really necessary for establishing the property.
    In practice, these techniques can even be used complementary to one another.

\section{Conclusion}
\label{sec:conc}

    We presented a technique to verify permutation invariance in DNNs. The novelty
    of our approach is a useful tie-class analysis, for forward propagation, and a
    scalable 2-polytope under-approximation method, for backward propagation. Our
    approach is sound (not just for permutation invariance properties, but for
    general safety properties too), efficient, and scalable. It is natural to
    wonder whether the approximately computed (reachable and safe) regions may be
    refined to eliminate spurious counterexamples, and continue the propagation
    till the property is proved or refuted.  Our approach is definitely amenable to
    a counterexample-guided refinement. In particular, the spurious counterexamples
    can guide us to split $\mathit{Relu}$ nodes (to refine over-approximations),
    and add additional safe regions (to refine under-approximations). This would
    require us to maintain sets of affine regions and convex polytopes at each
    layer, which is challenging but an interesting direction to pursue.

\bibliographystyle{abbrv}
\bibliography{bib}

\begin{thebibliography}{10}

\bibitem{DBLP:conf/kr/AkitundeLMP18}
M.~Akintunde, A.~Lomuscio, L.~Maganti, and E.~Pirovano.
\newblock Reachability analysis for neural agent-environment systems.
\newblock In M.~Thielscher, F.~Toni, and F.~Wolter, editors, {\em Principles of
  Knowledge Representation and Reasoning: Proceedings of the Sixteenth
  International Conference, {KR} 2018, Tempe, Arizona, 30 October - 2 November
  2018}, pages 184--193. {AAAI} Press, 2018.

\bibitem{DBLP:conf/tacas/MouraB08}
L.~M. de~Moura and N.~Bj{\o}rner.
\newblock {Z3: An Efficient SMT Solver}.
\newblock In {\em TACAS}, volume 4963 of {\em LNCS}, pages 337--340. Springer,
  2008.

\bibitem{DBLP:conf/nfm/DuttaJST18}
S.~Dutta, S.~Jha, S.~Sankaranarayanan, and A.~Tiwari.
\newblock Output range analysis for deep feedforward neural networks.
\newblock In A.~Dutle, C.~A. Mu{\~{n}}oz, and A.~Narkawicz, editors, {\em
  {NASA} Formal Methods - 10th International Symposium, {NFM} 2018, Newport
  News, VA, USA, April 17-19, 2018, Proceedings}, volume 10811 of {\em Lecture
  Notes in Computer Science}, pages 121--138. Springer, 2018.

\bibitem{DBLP:conf/atva/Ehlers17}
R.~Ehlers.
\newblock Formal verification of piece-wise linear feed-forward neural
  networks.
\newblock In D.~D'Souza and K.~N. Kumar, editors, {\em Automated Technology for
  Verification and Analysis - 15th International Symposium, {ATVA} 2017, Pune,
  India, October 3-6, 2017, Proceedings}, volume 10482 of {\em Lecture Notes in
  Computer Science}, pages 269--286. Springer, 2017.

\bibitem{dnnAbs-CAV2020}
Y.~Y. Elboher, J.~Gottschlich, and G.~Katz.
\newblock An abstraction-based framework for neural network verification.
\newblock In S.~K. Lahiri and C.~Wang, editors, {\em Computer Aided
  Verification}, pages 43--65, Cham, 2020. Springer International Publishing.

\bibitem{DBLP:conf/sigcomm/EliyahuKKS21}
T.~Eliyahu, Y.~Kazak, G.~Katz, and M.~Schapira.
\newblock Verifying learning-augmented systems.
\newblock In F.~A. Kuipers and M.~C. Caesar, editors, {\em {ACM} {SIGCOMM} 2021
  Conference, Virtual Event, USA, August 23-27, 2021}, pages 305--318. {ACM},
  2021.

\bibitem{absInt-ai2}
T.~Gehr, M.~Mirman, D.~Drachsler-Cohen, P.~Tsankov, S.~Chaudhuri, and
  M.~Vechev.
\newblock Ai2: Safety and robustness certification of neural networks with
  abstract interpretation.
\newblock In {\em 2018 IEEE Symposium on Security and Privacy (SP)}, Los
  Alamitos, CA, USA, may 2018. IEEE Computer Society.

\bibitem{DBLP:conf/cav/HuangKWW17}
X.~Huang, M.~Kwiatkowska, S.~Wang, and M.~Wu.
\newblock Safety verification of deep neural networks.
\newblock In R.~Majumdar and V.~Kuncak, editors, {\em Computer Aided
  Verification - 29th International Conference, {CAV} 2017, Heidelberg,
  Germany, July 24-28, 2017, Proceedings, Part {I}}, volume 10426 of {\em
  Lecture Notes in Computer Science}, pages 3--29. Springer, 2017.

\bibitem{DBLP:conf/atva/JacobyBK20}
Y.~Jacoby, C.~W. Barrett, and G.~Katz.
\newblock Verifying recurrent neural networks using invariant inference.
\newblock In D.~V. Hung and O.~Sokolsky, editors, {\em Automated Technology for
  Verification and Analysis - 18th International Symposium, {ATVA} 2020, Hanoi,
  Vietnam, October 19-23, 2020, Proceedings}, volume 12302 of {\em Lecture
  Notes in Computer Science}, pages 57--74. Springer, 2020.

\bibitem{DBLP:conf/cav/KatzBDJK17}
G.~Katz, C.~W. Barrett, D.~L. Dill, K.~Julian, and M.~J. Kochenderfer.
\newblock Reluplex: An efficient {SMT} solver for verifying deep neural
  networks.
\newblock In R.~Majumdar and V.~Kuncak, editors, {\em Computer Aided
  Verification - 29th International Conference, {CAV} 2017, Heidelberg,
  Germany, July 24-28, 2017, Proceedings, Part {I}}, volume 10426 of {\em
  Lecture Notes in Computer Science}, pages 97--117. Springer, 2017.

\bibitem{DBLP:conf/cav/KatzHIJLLSTWZDK19}
G.~Katz, D.~A. Huang, D.~Ibeling, K.~Julian, C.~Lazarus, R.~Lim, P.~Shah,
  S.~Thakoor, H.~Wu, A.~Zeljic, D.~L. Dill, M.~J. Kochenderfer, and C.~W.
  Barrett.
\newblock The marabou framework for verification and analysis of deep neural
  networks.
\newblock In I.~Dillig and S.~Tasiran, editors, {\em Computer Aided
  Verification - 31st International Conference, {CAV} 2019, New York City, NY,
  USA, July 15-18, 2019, Proceedings, Part {I}}, volume 11561 of {\em Lecture
  Notes in Computer Science}, pages 443--452. Springer, 2019.

\bibitem{DBLP:conf/sigcomm/KazakBKS19}
Y.~Kazak, C.~W. Barrett, G.~Katz, and M.~Schapira.
\newblock Verifying deep-rl-driven systems.
\newblock In {\em Proceedings of the 2019 Workshop on Network Meets {AI} {\&}
  ML, NetAI@SIGCOMM 2019, Beijing, China, August 23, 2019}, pages 83--89.
  {ACM}, 2019.

\bibitem{DBLP:conf/nips/PrabhakarA19}
P.~Prabhakar and Z.~R. Afzal.
\newblock Abstraction based output range analysis for neural networks.
\newblock In H.~M. Wallach, H.~Larochelle, A.~Beygelzimer,
  F.~d'Alch{\'{e}}{-}Buc, E.~B. Fox, and R.~Garnett, editors, {\em Advances in
  Neural Information Processing Systems 32: Annual Conference on Neural
  Information Processing Systems 2019, NeurIPS 2019, December 8-14, 2019,
  Vancouver, BC, Canada}, pages 15762--15772, 2019.

\bibitem{abs-refine-early-cav10}
L.~Pulina and A.~Tacchella.
\newblock An abstraction-refinement approach to verification of artificial
  neural networks.
\newblock In T.~Touili, B.~Cook, and P.~Jackson, editors, {\em Computer Aided
  Verification}, pages 243--257, Berlin, Heidelberg, 2010. Springer Berlin
  Heidelberg.

\bibitem{DBLP:conf/nips/SinghGMPV18}
G.~Singh, T.~Gehr, M.~Mirman, M.~P{\"{u}}schel, and M.~T. Vechev.
\newblock Fast and effective robustness certification.
\newblock In S.~Bengio, H.~M. Wallach, H.~Larochelle, K.~Grauman,
  N.~Cesa{-}Bianchi, and R.~Garnett, editors, {\em Advances in Neural
  Information Processing Systems 31: Annual Conference on Neural Information
  Processing Systems 2018, NeurIPS 2018, December 3-8, 2018, Montr{\'{e}}al,
  Canada}, pages 10825--10836, 2018.

\bibitem{DBLP:journals/pacmpl/SinghGPV19}
G.~Singh, T.~Gehr, M.~P{\"{u}}schel, and M.~T. Vechev.
\newblock An abstract domain for certifying neural networks.
\newblock {\em Proc. {ACM} Program. Lang.}, 3({POPL}):41:1--41:30, 2019.

\bibitem{syrenn}
M.~Sotoudeh and A.~V. Thakur.
\newblock Syrenn: {A} tool for analyzing deep neural networks.
\newblock In J.~F. Groote and K.~G. Larsen, editors, {\em Tools and Algorithms
  for the Construction and Analysis of Systems - 27th International Conference,
  {TACAS} 2021, Held as Part of the European Joint Conferences on Theory and
  Practice of Software, {ETAPS} 2021, Luxembourg City, Luxembourg, March 27 -
  April 1, 2021, Proceedings, Part {II}}, volume 12652 of {\em Lecture Notes in
  Computer Science}, pages 281--302. Springer, 2021.

\bibitem{Tjeng2019EvaluatingRO}
V.~Tjeng, K.~Y. Xiao, and R.~Tedrake.
\newblock Evaluating robustness of neural networks with mixed integer
  programming.
\newblock In {\em ICLR}, 2019.

\bibitem{DBLP:conf/tacas/WickerHK18}
M.~Wicker, X.~Huang, and M.~Kwiatkowska.
\newblock Feature-guided black-box safety testing of deep neural networks.
\newblock In D.~Beyer and M.~Huisman, editors, {\em Tools and Algorithms for
  the Construction and Analysis of Systems - 24th International Conference,
  {TACAS} 2018, Held as Part of the European Joint Conferences on Theory and
  Practice of Software, {ETAPS} 2018, Thessaloniki, Greece, April 14-20, 2018,
  Proceedings, Part {I}}, volume 10805 of {\em Lecture Notes in Computer
  Science}, pages 408--426. Springer, 2018.

\bibitem{DBLP:journals/tcs/WuWRHK20}
M.~Wu, M.~Wicker, W.~Ruan, X.~Huang, and M.~Kwiatkowska.
\newblock A game-based approximate verification of deep neural networks with
  provable guarantees.
\newblock {\em Theor. Comput. Sci.}, 807:298--329, 2020.

\bibitem{DBLP:journals/tnn/XiangTJ18}
W.~Xiang, H.~Tran, and T.~T. Johnson.
\newblock Output reachable set estimation and verification for multilayer
  neural networks.
\newblock {\em {IEEE} Trans. Neural Networks Learn. Syst.}, 29(11):5777--5783,
  2018.

\bibitem{DBLP:conf/ecai/ZhangSGGLN20}
H.~Zhang, M.~Shinn, A.~Gupta, A.~Gurfinkel, N.~Le, and N.~Narodytska.
\newblock Verification of recurrent neural networks for cognitive tasks via
  reachability analysis.
\newblock In G.~D. Giacomo, A.~Catal{\'{a}}, B.~Dilkina, M.~Milano, S.~Barro,
  A.~Bugar{\'{\i}}n, and J.~Lang, editors, {\em {ECAI} 2020 - 24th European
  Conference on Artificial Intelligence, 29 August-8 September 2020, Santiago
  de Compostela, Spain, August 29 - September 8, 2020 - Including 10th
  Conference on Prestigious Applications of Artificial Intelligence {(PAIS}
  2020)}, volume 325 of {\em Frontiers in Artificial Intelligence and
  Applications}, pages 1690--1697. {IOS} Press, 2020.

\end{thebibliography}

\newpage

\appendix
\section{Proofs}
\label{app:proofs}

    In the following proofs, we refer to the input affine region over-approximating the points before
    the $\mathit{Relu}$ by $A$, given by basis $B$ and center $c$.

    \noindent
    \textbf{Lemma \ref{l1}} Given $\overrightarrow{x} = \sum_i \alpha_i \overrightarrow{v_i} +
        \overrightarrow{c}$, we can write $\mathit{Relu}(\overrightarrow{x}) = \sum_{i,j} \alpha'^j_i
        \overrightarrow{v'^j_i} + \sum_j \beta_j \overrightarrow{c_j}$ where each $\alpha'^j_i$ is
        either $\alpha_i$ or is $0$, and each $\beta_j$ is either $0$ or $1$. Moreover, the components
        of $\mathit{Relu}(\overrightarrow{x})$ with indices in a tie class $j$ are $0$ if and only if
        $\alpha'^j_i$ and $\beta_j$ are $0$.

        \textit{Note:} Here $i$ is an index for the components of vectors, and varies between $0$ and
        $n-1$, where $n$ is the dimension of the underlying space. Say there are $t$ tie classes. Then,
        $j$ is an index for the tie class and varies between $0$ and $t-1$. Thus, there are a total $tn$
        terms in $\sum_{i,j} \alpha'^j_i \overrightarrow{v'^j_i}$ and $t$ in  $\sum_j \beta_j
        \overrightarrow{c_j}$.

        \begin{proof} We are given an $\overrightarrow{x} = \sum_i \alpha_i \overrightarrow{v_i}$. Since
            each pair of components of $\overrightarrow{x}$ in each tie class never have different
            signs, all the components of $\overrightarrow{x}$ in any given tie class will have the same
            sign.
        
            Say the components whose indices are in the tie classes $j_1, j_2, {\cdots} j_k$ have
            positive sign, the rest all have non-positive sign. Then, for any component of
            $\mathit{Relu}(\overrightarrow{x})$ whose index is in some $j_l$, $1\leq l\leq k$, their value will
            be the same as that of $\overrightarrow{x}$, as their corresponding component of
            $\overrightarrow{x}$ is positive. On the other hand, if the component's index is not in any
            $j_l$, it will be $0$ since the corresponding component of $\overrightarrow{x}$ is negative.
            
            Now consider the sum of vectors, $\sum_i \alpha_i \overrightarrow{v'^{j_l}_i} +
            \overrightarrow{c_{j_l}}$ for some $j_l$, $1\leq l\leq k$. There are $n$ many
            $\overrightarrow{v'^{j_l}_i}$s in this sum. For indices in $j_l$ the values of
            the components of $\overrightarrow{v'^{j_l}_i}$ and $\overrightarrow{v_i}$ are the same for
            the same $i$, and the component of $\overrightarrow{c_{j_l}}$ is the same as that of
            $\overrightarrow{c_j}$. So, for these indices, the component of $\sum_i \alpha_i
            \overrightarrow{v'^{j_l}_i} + \overrightarrow{c_{j_l}}$ is the same as that of
            $\overrightarrow{x}$. For any index outside $j_l$, the components of all the 
            index $\overrightarrow{v'^{j_l}_i}$ and that of $\overrightarrow{c_{j_l}}$ is $0$. So, for
            these indices, the component of $\sum_i \alpha_i \overrightarrow{v'^{j_l}_i} +
            \overrightarrow{c_{j_l}}$ is $0$.
            
            Now, consider the following sum of terms of the above form, with $k$ many terms. This is a
            sum that involves $nk$ many $\overrightarrow{v'^j_i}$, while the rest of the $n(t-k)$ of
            these $\overrightarrow{v'^j_i}$ may be considered to have a zero component here. Similarly,
            there are $k$ $\overrightarrow{c_j}$s, and $t-k$ $\overrightarrow{c_j}$s may be considered
            to have $0$ components.
            
            \begin{equation*}
            \begin{aligned}
                \sum_i \alpha_i \overrightarrow{v'^{j_1}_i} + \overrightarrow{c_{j_1}} + \sum_i \alpha_i
                \overrightarrow{v'^{j_2}_i} + \overrightarrow{c_{j_2}} + {\cdots} + \sum_i \alpha_i
                \overrightarrow{v'^{j_k}_i} + \overrightarrow{c_{j_k}}
            \end{aligned}
            \end{equation*}

            For any index in some $j_l$, $1\leq l\leq k$, the component of $\sum_i \alpha_i
            \overrightarrow{v'^{j_l}_i} + \overrightarrow{c_{j_l}}$ has the same value as
            $\overrightarrow{x}$, while the components of all the other terms are $0$. Thus, the
            component of the sum is the same as that of $\overrightarrow{x}$. For any index not in any
            $j_l$, the components of all the terms are $0$, so the component of the sum is $0$. Hence,
            we have that the sum is $\mathit{Relu}(\overrightarrow{x})$.
            
            Thus, we can set $\alpha'^{j_l}_i$ to $\alpha_i$ for all $j_l$, and for other tie classes,
            we can set $\alpha^j_i$ to $0$. Similarly, we can set $\beta_{j_l}$ to $1$ for all $j_l$,
            and other tie classes we set $\beta_j$ to 0. This gives us an expression for
            $\mathit{Relu}(x)$ of the required form. Also, we see that the components of
            $\mathit{Relu}(\overrightarrow{x})$ with indices in a tie class $j$ are $0$ if and only if
            $\alpha'^j_i$ and $\beta_j$ are $0$.  \qed
        \end{proof}

        To prove theorem \ref{t1}, we prove two intermediate lemmas.

    \noindent
        \begin{lemma} \label{l2} For any given tie class $j$, the components of $\overrightarrow{c}$ in
            the tie class are either all positive, or all non-positive.
        \end{lemma}
        
        \begin{proof} This comes directly from the definition of tie class. We
            notice that if we set all $\alpha_i$ to $0$, we get that $\overrightarrow{c}$ is a vector in
            $A$.  Then, by definition, it's components that are in the same tie class will
            all be positive, or all be negative.\qed
        \end{proof}

    \noindent
        \begin{lemma} \label{l3} Given two vectors $\overrightarrow{a}$ and $\overrightarrow{b}$ and a
            tie class $j_0$ so that:
            \begin{enumerate}
                \item $\overrightarrow{a}$ and $\overrightarrow{a} + \overrightarrow{b}$ are in $A$
                \item the nonzero components of $\overrightarrow{a}$ all have indices in $j_0$
                \item for any index $i$, if the $i$ component of $\overrightarrow{b}$, $b_i \neq 0$,
                    then $i \in j_0$.
                \item the components of $\overrightarrow{a}$ and $\overrightarrow{a} + \overrightarrow{b}$
                    with indices in $j_0$ have different signs
            \end{enumerate}
            then there is a $0 \leq \gamma \leq 1$ so that $\overrightarrow{a} = - \gamma \overrightarrow{b}$.
        \end{lemma}
        
        \begin{proof} Let $\overrightarrow{d} = \overrightarrow{a} + \overrightarrow{b}$. As
            $\overrightarrow{a}$ and $\overrightarrow{d}$ are in $A$, we have:

            \begin{equation*}
            \begin{aligned}
                \overrightarrow{a} = \sum_i \alpha^a_i \overrightarrow{v_i} + \overrightarrow{c} \\
                \overrightarrow{d} = \sum_i \alpha^d_i \overrightarrow{v_i} + \overrightarrow{c} \\
            \end{aligned}
            \end{equation*}\linebreak
            
            For some $-1 \leq \alpha^a_i, \alpha^d_i \leq 1$. Now, for any real $0 < \gamma < 1$
            
            \begin{equation*}
            \begin{aligned}
                \overrightarrow{a} + \gamma \overrightarrow{b}
                &= (1 - \gamma) \overrightarrow{a} + \gamma \overrightarrow{d} \\
                &= \sum_i ( (1 - \gamma) \alpha^a_i + \gamma \alpha^d_i ) \overrightarrow{v_i} + \overrightarrow{c} \\
            \end{aligned}
            \end{equation*}\linebreak
            
            As $-1 \leq (1 - \gamma) \alpha^a_i + \gamma \alpha^d_i \leq 1$, $\overrightarrow{a} +
            \gamma \overrightarrow{b}$ lies in $A$. All nonzero components of
            $\overrightarrow{a}$ and $\overrightarrow{b}$ have indices that lie in the same tie class
            $j_0$, and so all nonzero components of $\overrightarrow{a} + \gamma \overrightarrow{b}$
            have indices in $j_0$ as well. By definition of tie class, for any $0 < \gamma < 1$, all
            these components must have the same sign. So, all components of $\overrightarrow{a} +
            \gamma \overrightarrow{b}$ with indices in $j_0$ have the same sign.
            
            Now, consider the function that takes $\gamma$ to a component of $\overrightarrow{a} +
            \gamma \overrightarrow{b}$ with an index belonging to $j_0$. This is a real valued function
            on $(0, 1)$. At $\gamma = 0$, $\overrightarrow{a} + \gamma \overrightarrow{b} =
            \overrightarrow{a}$, and at $\gamma = 1$, $\overrightarrow{a} + \gamma
            \overrightarrow{b} = \overrightarrow{a} + \overrightarrow{b}$. So the value of the function
            at $0$ and $1$ must have opposite signs. Then, there is some $0 \leq \gamma_0 \leq 1$ for
            which the value of the function is $0$. But, since all the components of $\overrightarrow{a}
            + \gamma_0 \overrightarrow{b}$ with indices in $j_0$ have the same sign, all these
            components must be zero. All other components of $\overrightarrow{a} + \gamma_0
            \overrightarrow{b}$ is zero as well. So, we have:
            
            \begin{equation*}
            \begin{aligned}
                \overrightarrow{a} + \gamma \overrightarrow{b} = 0 \;{\Rightarrow}\;
                \overrightarrow{a} = - \gamma \overrightarrow{b}
            \end{aligned}
            \end{equation*}\linebreak \qed
        \end{proof}

    \noindent
    \textbf{Theorem \ref{t1}} Given a $\overrightarrow{x} = \sum_i \alpha_i \overrightarrow{v_i}
        + \overrightarrow{c}, |\overrightarrow{\alpha_i}| \leq 1$ in $A$, there are scalars
        $\alpha'^j_i$ so that:
        \begin{enumerate}
            \item $\mathit{Relu}(\overrightarrow{x}) = \sum_{i,j} \alpha'^j_i \overrightarrow{v'^j_i} +
                \mathit{Relu}(\overrightarrow{c})$
            \item $|\overrightarrow{\alpha'^j_i}| \leq 1$ for all $i$ and $j$.
        \end{enumerate}

        \begin{proof} Given any $\overrightarrow{x}$ of the given form, we firstly apply lemma \ref{l1} to get the
            following expression for $\mathit{Relu}(\overrightarrow{x})$:
            
            \begin{equation*}
            \begin{aligned}
                \mathit{Relu}(\overrightarrow{x}) = \sum_{i,j} \alpha'^j_i \overrightarrow{v'^j_i} + \sum_j
                \beta_j \overrightarrow{c_j}
            \end{aligned}
            \end{equation*}
            
            Where each $\alpha'^j_i$ is either $0$ or $\alpha_i$, and each $\beta_j$ is $0$ or $1$.
            
            Now, we know from lemma \ref{l2} that for each tie class, the components of
            $\overrightarrow{c}$ are all positive, or all negative. So, each $\overrightarrow{c_j}$ is
            either all positive, or all negative. Thus, we have:
            
            \begin{equation*}
            \begin{aligned}
                \mathit{Relu}(\overrightarrow{c}) = \sum_j \beta'_j \overrightarrow{c_j}
            \end{aligned}
            \end{equation*}
            
            Where $\beta'_j$ is $1$ if $\overrightarrow{c_j}$ has all positive components, and $0$ otherwise.
            
            For each tie class $j$, we wish to replace $\sum_j \beta_j \overrightarrow{c_j}$ in
            $\mathit{Relu}(\overrightarrow{x})$ with $\mathit{Relu}(\overrightarrow{c})$ by adjusting
            the values of $\alpha'^j_i$. So, it suffices to find $\alpha''^j_i$ so that the following
            holds:
           
            \begin{equation*}
            \begin{aligned}
                \sum_i \alpha'^j_i \overrightarrow{v'^j_i} + \beta_j \overrightarrow{c_j} = 
                \sum_i \alpha''^j_i \overrightarrow{v'^j_i} + \beta'_j \overrightarrow{c_j} =
                \sum_i \alpha''^j_i \overrightarrow{v'^j_i} + \mathit{Relu}(\overrightarrow{c})
            \end{aligned}
            \end{equation*}

            To do this adjustment, we look at four cases, depending on the sign of components of
            $\overrightarrow{x}$ in $j$, and the sign of components of $\overrightarrow{c_j}$:
            
            \textbf{Case 1:}\textit{Components of $\overrightarrow{x}$ in $j$ and components of
            $\overrightarrow{c_j}$ are both positive.} In this case, by lemma \ref{l1}, $\beta_j$ is
            $1$, and as components of $\overrightarrow{c_j}$ are positive $\beta'_j$ is also $1$. So, we
            can set $\alpha''^j_i$ to $\alpha'^j_i$.

            \textbf{Case 2:}\textit{Components of $\overrightarrow{x}$ in $j$ and components of
            $\overrightarrow{c_j}$ are both negative or 0.} In this case, since
            $\mathit{Relu}(\overrightarrow{x}) = 0$ we can simply set $\alpha''^j_i = 0$.
            
            \textbf{Case 3:}\textit{Components of $\overrightarrow{x}$ in $j$ are positive, but
            components of $\overrightarrow{c_j}$ are negative or 0.} In this case, $\beta'_j$ is $0$, and by
            lemma \ref{l1} $\beta_j$ is $1$, and the components of $\overrightarrow{x}$ with indices in
            the tie class $j$ are same as that of $\sum_i \alpha_i \overrightarrow{v'^j_i} +
            \overrightarrow{c_j}$, with $\alpha'^j_i = \alpha_i$. Now, if we set $\overrightarrow{a} =
            \overrightarrow{c_j}$ and $\overrightarrow{b} = \sum_i \alpha_i \overrightarrow{v'^j_i}$, we
            can use lemma \ref{l3} to get a $\gamma$ so that:
            
            \begin{equation*}
            \begin{aligned}
                \; & \overrightarrow{a} = -\gamma \overrightarrow{b} \\
                \;{\Rightarrow}\; & \overrightarrow{c_j} = -\gamma \sum_i \alpha_i
                \overrightarrow{v'^j_i} \\
                \;{\Rightarrow}\; & \sum_i \alpha_i \overrightarrow{v'^j_i} + \beta_j
                \overrightarrow{c_j} = \sum_i (1-\gamma) \alpha_i \overrightarrow{v'^j_i} + \beta'_j
                \overrightarrow{c_j} \\
            \end{aligned}
            \end{equation*}
            
            Now, as $0 \leq \gamma \leq 1$, we have $0 \leq 1-\gamma \leq 1$, and so if we set $\alpha''^j_i =
            (1-\gamma)\alpha_i$, we have $ | \alpha''^j_i | \leq | \alpha_i | \leq 1$. Thus, we have found the
            required $\alpha''^j_i$.

            \textbf{Case 4:}\textit{Components of $\overrightarrow{x}$ in $j$ are non-positive, but
            components of $\overrightarrow{c_j}$ are positive.} Similarly to case 3, $\beta'_j$ is $1$,
            and lemma \ref{l1} gives us $\beta_j$ and $\alpha'^j_i$ are all $0$. Again we set $\overrightarrow{a} =
            \overrightarrow{c_j}$ and $\overrightarrow{b} = \sum_i \alpha_i \overrightarrow{v'^j_i}$ and
            use lemma \ref{l3} to get a $\gamma$ so that: 
            
            \begin{equation*}
            \begin{aligned}
                \; & \overrightarrow{a} = -\gamma \overrightarrow{b} \\
                \;{\Rightarrow}\; & \overrightarrow{c_j} = -\gamma \sum_i \alpha_i
                \overrightarrow{v'^j_i} \\
                \;{\Rightarrow}\; & \sum_i \alpha_i v'^j_i + \beta_j c_j = \sum_i \gamma \alpha_i v'^j_i + c_j
            \end{aligned}
            \end{equation*}
            
            Again, we can set $\alpha''^j_i = \gamma \alpha_i$, and as $0 \leq \gamma \leq 1$, the bounds are
            satisfied.
            
            Thus, given an $\overrightarrow{x}$, for each tie class $j$ we can always find the required
            $\alpha''^j_i$, and we are done. \qed
        \end{proof}
        
    \noindent
    \textbf{Theorem \ref{t2}} Two indices $i$ and $j$ are in the same tie class if
        and only if one of the following is true:
        \begin{enumerate}
            \item The $i$ and $j$ components of $\overrightarrow{x}$ are always both positive.
            \item The $i$ and $j$ components of $\overrightarrow{x}$ are always both negative.
            \item The vector formed by the $i$ and $j$ components of the $\overrightarrow{v_k}$ and
                $\overrightarrow{c}$ are parallel. In other words, if $v^l_k$ is the $l$-component of
                $\overrightarrow{v_k}$, and $c^l$ is the $l$ component of $\overrightarrow{c}$, then $[
                v^i_1, v^i_2, {\cdots} c^i ] = k[ v^j_1, v^j_2, {\cdots} c^j ]$ for some real $k > 0$. 
        \end{enumerate}
        
        \begin{proof} \textbf{Forward direction:} Say $i$ and $j$ are in the same tie class and that $1$
        and $2$ do not hold. We show that $3$ must hold.
        
        Since $1$ and $2$ does not hold, we can take a $\overrightarrow{x_1}$ in $A$ where both $i$ and
        $j$ components are positive, and a $\overrightarrow{x_2}$ where both are negative. We use an
        argument similar to the proof of Lemma \ref{l3}. 
        
        Consider the line segment joining the two points. The points on this line segment have the form
        $\lambda \overrightarrow{x_1} + (1 - \lambda) \overrightarrow{x_2}$, $0 \leq \lambda \leq 1$.
        Since $A$ is linear, all the points on this line segment are in $A$. Consider the $i$ coordinate
        of the points on this line. It is positive when $\lambda = 0$, and negative when $\lambda = 1$.
        So, for some value of $\lambda_0$, the $i$ coordinate of $\lambda_0 x_1 + (1 - \lambda_0) x_2$l
        is $0$. Since $i$ and $j$ are in the same tie class, $j$ component of this point is also $0$.
        So, if the $i$ and $j$ components of $x_1$ is $x^i_1$ and $x^j_1$, and that of $x_2$ be $x^i_2$
        and $x^j_2$ respectively, we have:
        
        \begin{equation*}
        \begin{aligned}
            \; & \lambda x^i_1 + (1 - \lambda) x^i_2 = 0 \\
            \;{\Rightarrow}\; & \frac{x^i_1}{x^i_2} = \frac{\lambda-1}{\lambda}\\
            \mathit{and},\; & \lambda x^j_1 + (1 - \lambda) x^j_2 = 0 \\
            \;{\Rightarrow}\; & \frac{x^j_1}{x^j_2} = \frac{\lambda-1}{\lambda} = \frac{x^i_1}{x^i_2}\\
            \;{\Rightarrow}\; & \frac{x^i_1}{x^j_1} = \frac{x^i_2}{x^j_2} = k\\
        \end{aligned}
        \end{equation*}
        
        Now, if we pick any $\overrightarrow{x} \neq 0$ in $A$, if it's $i$-component is positive, it's
        $j$ component must also be positive. Then, we can replace $\overrightarrow{x_1}$ with
        $\overrightarrow{x}$ in the above argument to derive that the ratio of the $i$ component to
        the $j$ component of $\overrightarrow{x}$ must be $k$. Similarly, if $i$-component of
        $\overrightarrow{x}$ is negative, we can replace $x_2$ with it. So, for any $\overrightarrow{x}$
        in $A$, the $i$ component of $\overrightarrow{x}$ is $k$ times the $j$ component. Since $i$ and
        $j$ are in the same tie class, $k$ must be positive.
        
        Let the vector formed by the $i$ components of $\overrightarrow{v_i}$ and $\overrightarrow{c}$ be,
        $\overrightarrow{u^i}$ and that from the $j$ components be $\overrightarrow{u^j}$. For each
        $\overrightarrow{x} = \sum_i \alpha_i \overrightarrow{v_i} + \overrightarrow{c}$, we let
        $\overrightarrow{\alpha}$ be a vector whose last component is $1$, and the $i$ component is
        $\alpha_i$, that is, $\overrightarrow{\alpha} = [ \alpha_1 \alpha_2 {\cdots} 1 ]$. Then, $x^i =
        \overrightarrow{\alpha}.\overrightarrow{u^i}$, and $x^j =
        \overrightarrow{\alpha}.\overrightarrow{u^j}$, where $x^i$ and $x^j$ are the $i$ and $j$
        components of $\overrightarrow{x}$. Quantifying over all $\overrightarrow{x}$, we
        have:
        
        \begin{equation*}
        \begin{aligned}
            \; & \forall \overrightarrow{\alpha},\; \overrightarrow{\alpha}.\overrightarrow{u^i} =
            k\overrightarrow{\alpha}.\overrightarrow{u^j} \\ 
            \;{\Rightarrow}\; & \overrightarrow{u^i} = k \overrightarrow{u^j}
        \end{aligned}
        \end{equation*}
        
        So, $u^i$ and $u^j$ are parallel and we are done with the proof of the forward direction.
        
        \textbf{Backward direction:} It is clear that if $1$ or $2$ hold, $i$ and $j$ must be in the same
        tie class. If $3$ holds, we have, borrowing the notation from the proof of the forward
        direction, some $k > 0$ for which:
        
        \begin{equation*}
        \begin{aligned}
            \; & \overrightarrow{u^i} = k \overrightarrow{u^j} \\
            \;{\Rightarrow}\; & \forall \overrightarrow{\alpha},\; \overrightarrow{\alpha}.\overrightarrow{u^i} =
            k\overrightarrow{\alpha}.\overrightarrow{u^j} \\ 
        \end{aligned}
        \end{equation*}
        
        So, for all $x$ in $A$, $x^i = k x^j$, $k > 0$, so $x^i$ and $x^j$ have the same
        sign. Thus, $i$ and $j$ are in the same tie class. This completes the proof of the backward
        direction.\qed
        \end{proof}
        
    \noindent
    \textbf{Lemma \ref{l5}} The maximum and minimum values of $ \sum_i \alpha_i v_i $, for real
        $\alpha_i$, fixed real $v_i$, constrained by $| \alpha_i | \leq 1$, are $\sum_i |v_i|$ and
        $-\sum_i |v_i|$ respectively.

        \begin{proof} Say the maximum is achieved by some set of $\alpha_i$. Then, if for some $i$
            $\alpha_i$ and $v_i$ have opposite signs, then inverting the sign of $\alpha_i$ gives us a
            larger value. So, $\alpha_i$ must have the same sign as $v_i$, and $\alpha_i v_i = | \alpha_i
            v_i | = | \alpha_i | | v_i |$. Now, if we can increase the absolute value of $\alpha_i$, the
            value of the sum must increase, so we must have $| \alpha_i | = 1$. Thus, the maximum value is
            given by $\sum_i |v_i|$. If the minimum is less than $-\sum_i |v_i|$, inverting it's sign
            will give a maximum greater than $\sum_i |v_i|$. The value of $-\sum_i |v_i|$ can be
            attained by reversing the signs of the $\alpha_i$ that gives us the maximum. So, the minimum
            must be $-\sum_i |v_i|$.\qed
        \end{proof}
    
\section{Details of Backpropagation around 0}
\label{app:pb0_details}

    As stated in section \ref{subsec:backprop}, there are two steps for backpropagation around 0. The
    first steps finds one optimal $\overrightarrow{\eta_i}$ satisfying inequality in the input
    convex polytope $P$. The second step takes the component wise minimum of the
    $\overrightarrow{\eta_i}$ to get the required $\overrightarrow{\eta}$. Splitting the process of
    finding the $\overrightarrow{\eta}$ into two steps allows us to reduce this to solving sets of linear
    equations.

    \textbf{Optimizing for a single inequality:}
        
        In this step, for each inequality $\overrightarrow{x} . \overrightarrow{w} \leq u$ in $P$,
        we find an optimal $\overrightarrow{\eta}$ satisfying $\overrightarrow{\eta} .
        \overrightarrow{w} \leq u$. If $P$ is given by $\overrightarrow{x} L \leq
        \overrightarrow{u}$, then $\overrightarrow{w}$ a row of $L$, and $u$ is the corresponding
        component of $\overrightarrow{u}$.

        First, we show that we can assume without loss of generality that $w^0_i = 1$, where $w^0_i$ is
        the $0$ component of $\overrightarrow{w}$. We can do this since $\overrightarrow{w} \neq 0$,
        so at least one component is non-zero, and we can call it $w^0_i$. Then, we can scale all other
        components of $\overrightarrow{w}$ and $u$ by $w^0_i$ and set $w^0_i=1$ without changing the
        inequality. 

        If $\overrightarrow{\eta}.\overrightarrow{w} < u$, we can increase at least one component of
        $\overrightarrow{\eta}$ keeping the others unchanged, and so we can increase the product. So,
        an optimal $\overrightarrow{\eta}$ must satisfy $\overrightarrow{\eta}.\overrightarrow{w} = u$.
        Now, we can substitute $\eta_0$ from $\overrightarrow{\eta}.\overrightarrow{w} = u$ into
        $\prod_i \eta_i$ to get:

        \begin{equation*}
        \begin{aligned}
            \prod^{n-1}_{i=0} \eta_i = 
            \left(u - \sum_{i=1}^{n-1} \eta_i v_i \right) \prod_{i=1}^{n-1} \eta_i = 
        \end{aligned}
        \end{equation*}

        Now, we use the following standard fact: for any real valued function with $m$ real valued inputs
        $f(x_1, x_2, {\cdots}, x_n)$, at an input where $f$ is maximum the partial derivative of $f$
        with respect to each $x_i$ must be $0$. We define $f$ as:

        \begin{equation*}
        \begin{aligned}
            f(\eta_1, \eta_2, {\cdots} \eta_{n-1})
            = \left(u - \sum_{i=1}^{n-1} \eta_i v_i \right) \prod_{i=1}^{n-1} \eta_i
        \end{aligned}
        \end{equation*}\linebreak

        Then we can take the partial derivative of $f$ with respect to each $\eta_i$ and equate it to
        $0$ to get conditions that an optimal $\eta_i$ must satisfy:

        \begin{equation*}
        \begin{aligned}
            \;{\Rightarrow}\; & \frac{\partial f}{\partial \eta_j} = v_j \prod_{i=1}^{n-1} \eta_i + 
                \left(u - \sum_{i=1}^{n-1} \eta_i v_i \right) \prod_{i=1}^{j-1} \eta_i \prod_{i=j+1}^{n-1} \eta_i
                = 0\\
            \;{\Rightarrow}\; & v_j \eta_j + u - \sum_{i=1}^{n-1} \eta_i v_i = 0\\
        \end{aligned}
        \end{equation*}

        Here we have assumed that no $\eta_i$ is never $0$, which we can do since if any $\eta_i$ is
        $0$, the product becomes $0$.

        Thus, we have reduced finding the optimal $\overrightarrow{\eta_i}$ to solving a set of linear
        equations, one for each $j$. Since $0$ lies in the interior of the polytope, and our equation is
        a constraint in the polytope, we have $u > 0$. So, we can always solve the above equations to
        get $\eta_i > 0$.

    \textbf{Finding an $\overrightarrow{\eta}$ satisfying all inequalities:}
        We note that if $\overrightarrow{\eta}.\overrightarrow{w} \leq u$ holds then $0 \leq
        \overrightarrow{x} \leq \overrightarrow{\eta} \;{\Rightarrow}\;
        \overrightarrow{x}.\overrightarrow{w} \leq u$ holds, since all involved vectors are positive.
        Then, if we find some $\overrightarrow{\eta}$ so that
        $\overrightarrow{\eta}.\overrightarrow{w_k} \leq u_k$ holds for all inequalities
        $\overrightarrow{\eta}.\overrightarrow{w_k} \leq u_k$ in the input polytope, we have:
        
        \begin{equation*}
        \begin{aligned}
            \; & \; &\forall~k: \; 0 \leq \overrightarrow{x} \leq \overrightarrow{\eta} 
            \;{\Rightarrow}\; & \overrightarrow{x}.\overrightarrow{w_k} \leq u_k\\
            \;{\Rightarrow}\; & \; &
            \overrightarrow{x} 0 \leq \overrightarrow{x} \leq \overrightarrow{\eta}
            \;{\Rightarrow}\; & \overrightarrow{x} L \leq \overrightarrow{u}
        \end{aligned}
        \end{equation*}\linebreak

        To find such an $\overrightarrow{\eta}$, we repeat the above process for each inequality in
        $\overrightarrow{x}.\overrightarrow{w_k} \leq u_k$, and for each we get an set of $\eta^k_i$,
        one for each inequality. Then, if we take $\eta_i$ as the minimum of $\eta^k_i$ over all $k$,
        the obtained $\overrightarrow{\eta}$ will satisfy all inequalities in $\overrightarrow{x} L \leq
        \overrightarrow{u}$.

\end{document}